# Reconstruction of a Long-term spatially Contiguous Solar-Induced Fluorescence (LCSIF) over 1982-2022

Jianing Fang[a*], Xu Lian[a], Youngryel Ryu[b,c,d], Sungchan Jeong[b,c], Chongya Jiang[e], Pierre Gentine[a]

a. Department of Earth and Environmental Engineering, Columbia University, New York 10027, NY, USA

b. Interdisciplinary Program in Landscape Agriculture, Seoul National University, Seoul, Republic of Korea

c. Integrated Major in Smart City Global Convergence, Seoul National University, Seoul, Republic of Korea

d. Department of Landscape Architecture and Rural Systems Engineering, Seoul National University, Seoul, Republic of Korea

e. Department of Natural Resources and Environmental Sciences, University of Illinois at Urbana-Champaign

* Corresponding author. Email: jf3423@columbia.edu

## Highlights

Globally reconstructed SIF is extended back to 1982 with calibrated AVHRR reflectance

Red and NIR bands are sufficient for neural networks to capture the main SIF patterns

Site-level validation shows reconstructed SIF better approximates GPP than most VIs

## Abstract

Satellite-observed solar-induced chlorophyll fluorescence (SIF) is a powerful proxy for diagnosing the photosynthetic characteristics of terrestrial ecosystems. Despite the increasing spatial and temporal resolutions of these satellite retrievals, records of SIF are primarily limited to the recent decade, impeding their application in detecting long-term dynamics of ecosystem function and structure. In this study, we leverage the two surface reflectance bands (red and near-infrared) available both from Advanced Very High-Resolution Radiometer (AVHRR, 1982-2022) and MODerate-resolution Imaging Spectroradiometer (MODIS, 2001-2022). Importantly, we calibrate and orbit-correct the AVHRR bands against their MODIS counterparts during their overlapping period. Using the long-term bias-corrected reflectance data, a neural network is then built to reproduce the Orbiting Carbon Observatory-2 SIF using AVHRR and MODIS, and used to map SIF globally over the entire 1982-2022 period. Compared with the previous MODIS-based CSIF product relying on four reflectance bands, our two-band-based product has similar skill but can be advantageously extended to the bias-corrected AVHRR period. Further comparison with three widely used vegetation indices (NDVI, kNDVI, NIRv; all based empirically on red and near-infrared bands) shows a higher or comparable correlation of LCSIF with satellite SIF and site-level GPP estimates across vegetation types, ensuring a greater capacity of LCSIF for representing terrestrial photosynthesis. Globally, LCSIF-AVHRR shows an accelerating upward trend since 1982, with an average rate of 0.0025 mW $m^{-2}$ $nm^{-1}$ $sr^{-1}$ per decade during 1982-2000 and 0.0038 mW $m^{-2}$ $nm^{-1}$ $sr^{-1}$ per decade during 2001-2022. Previous greening hotspots of India and China since 2000 are also identified as hotspots for enhanced photosynthesis. Our LCSIF data provide opportunities to better understand the long-term dynamics of ecosystem photosynthesis and their underlying driving processes.

## 1 Introduction

Photosynthesis by terrestrial vegetation is the largest driver of the global biogeochemical cycles, absorbing a considerable fraction of atmospheric $CO_2$ annually and slowing down anthropogenic global warming (Friedlingstein et al., 2022). Obtaining spatiotemporally contiguous proxies of gross primary productivity (GPP) is crucial for understanding the climatic benefits of carbon cycle feedbacks, and for accurately forecasting forestry and agricultural yields (Badgley et al., 2017). Although the biochemistry of photosynthesis is well-characterized at the leaf level (Farquhar et al., 1980), this carbon flux cannot be monitored directly via large-scale approaches such as spaceborne remote sensing. To date, major uncertainties still exist in the regional and global estimates of carbon uptake by terrestrial ecosystems (Keenan and Williams, 2018) and in the long-term retrieval of those carbon fluxes, limiting our understanding of the drivers of interannual and decadal variability in the biogeochemical cycle.

Traditionally, global estimates of photosynthesis can be categorized into one of the three following approaches: (1) empirical light-use efficiency models (Monteith, 1972; Running et al., 2004; Zhang et al., 2017); (2) simplified process-based radiative transfer and biochemical models (Jiang and Ryu, 2016); (3) machine learning-based upscaling of site-level eddy-covariance flux measurements (Tramontana et al., 2016); (4) Data-assimilation approaches to constrain model posterior predictive fluxes with site-level and remote sensing observations (Luo et al., 2011). Empirical light-use efficiency models use a light-use efficiency (LUE) assumption that approximates GPP as a product of the incident photosynthetically active radiation (PAR), the fraction of absorbed PAR (fPAR), and the LUE. These models use meteorological information to adjust the LUE and to generate GPP predictions at moderately fine temporal and spatial resolutions (Zhang et al., 2017). However, differences in how meteorological drivers are included in the formulations of LUE models result in disparate performance for simulating the temporal patterns of GPP on a regional scale or estimating the responses of GPP to drought events (Chen et al., 2017; Stocker et al., 2019). Process-based models can integrate the knowledge of the soil-vegetation-atmosphere continuum and resolve detailed process representations (Farquhar et al., 1980; Williams et al., 1996), but the model's complexity can make computations at high spatial and temporal resolutions prohibitively expensive (Jiang and Ryu, 2016) and model tuning very challenging (Luo et al., 2012). While developments in machine-learning models driven by Earth observations have vastly advanced our ability to upscale in-situ eddy-covariance carbon flux measurements globally (Tramontana et al., 2016), systematic biases remain in simulating photosynthesis, especially in terms of interannual variabilities and the long-term trends of carbon uptake (Jung et al., 2020). The biases of the data-driven machine-learning upscaling models are especially large in the wet tropics, limiting our ability to infer the trends and interannual variabilities of the global terrestrial carbon cycle (Jung et al., 2020). Data-assimilation approaches adopt a Bayesian approach to constrain prior model parameters with observations with the goal of deriving better predictions and principled uncertainty quantifications. Prominent examples data assimilation models include but are not limited to ORCHIDAS (Bacour et al., 2023), CARDAMOM (Bloom et al., 2016; Bloom and Williams, 2015; Quetin et al., 2023; Yang et al., 2022), and CLM-DART (Fox et al., 2022). Nevertheless, the quality of posterior predictive fluxes still depend on the information content of the observations assimilated (Smallman et al., 2017), the identifiability of the model parameters (Bloom and Williams, 2015), and potential structural uncertainties in the model formulation (Famiglietti et al., 2021).

The past decade has witnessed tremendous successes with the use of satellite-observed passive solar-induced chlorophyll fluorescence as a proxy of gross primary productivity. Chlorophyll fluorescence is the emission of red and far-red photons from the excited states of Chl a molecules that provides an alternative energy dissipation pathway in addition to photochemical and nonphotochemical quenching (Gu et al., 2019; Porcar-Castell et al., 2014). Both leaf-level carbon assimilation ($A_{leaf}$) and chlorophyll fluorescence (ChlF) can be written using LUE-based expressions (Monteith, 1972):

$$ChlF = PAR \times fPAR_{chl} \times \Theta_F \quad (1)$$

$$A_{leaf} = PAR \times fPAR_{chl} \times \Theta_{P,\ leaf} \quad (2)$$

where $\Theta_F$ and $\Theta_{P,\ leaf}$ represent the quantum yield for ChlF emission and leaf-level photochemistry respectively. PAR stands for photosynthetically active radiation, while the $fPAR_{chl}$ term represents the fraction of PAR absorbed by chlorophyll, which is a function of the canopy structure or the biophysical properties of vegetation, and can often be approximated using satellite-observed optical vegetation indices (VIs) or visible and near-infrared reflectance channels (Gentine and Alemohammad, 2018; Zeng et al., 2022). When measured at ecosystem scale by spaceborne sensors, satellite SIF ($SIF_{sat}$, Equation 3) can be approximated with a similar equation with additional terms considering the fraction of SIF escaping from the canopy ($f_{esc}$) and atmospheric transmittance ($\tau_{atm}$) (Joiner et al., 2014).

$$SIF_{sat} = PAR \times fPAR_{chl} \times \Theta_F \times f_{esc} \times \tau_{atm} \quad (3)$$

$$GPP = PAR \times fPAR_{chl} \times \Theta_{P,\ canopy} \quad (4)$$

On a canopy to ecosystem scale, the total amount of organic carbon fixed by all green plants is captured by the notion of gross primary productivity (GPP). If we take the big-leaf approach which assumes a linear scaling between $A_{leaf}$ and GPP (Friend, 2001; Sellers et al., 1992), then

GPP can also be expressed as a light-use efficiency model (Equation 4), where $\Theta_{P,\ canopy}$ is the canopy-scale photochemistry quantum yield. It is considered that the one of main advantages of using $SIF_{sat}$ as a proxy of GPP compared with traditional reflectance-based VIs derives from the physiological coupling between the $\Theta_F$ and $\Theta_P$, so that $SIF_{sat}$ is more sensitive to abiotic stress than reflectance-based VIs (Zhang et al., 2018a). Although the instantaneous SIF-GPP relationship can be non-linear — with GPP saturating at high solar irradiance and SIF largely scaling linearly with light (Gu et al., 2019; Kim et al., 2021; Zhang et al., 2018c) — studies have found a robust near linear SIF-GPP relationship from daily to monthly scale using the OCO-2 SIF product (Wood et al., 2017; Zhang et al., 2018c). These findings have encouraged the use of satellite-based SIF for many applications such as tracking dynamics of GPP (Pierrat et al., 2022), phenological timing (Zhang et al., 2020) and crop yield (Peng et al., 2020), and for carbon flux partitioning (Zhan et al., 2022), drought monitoring (Sun et al., 2015), and assimilation into global biogeochemical models (Quetin et al., 2023).

Nevertheless, quantitatively connecting GPP to satellite SIF is hindered by the spatiotemporal discontinuities in the current generation of satellite SIF retrievals (Porcar-Castell et al., 2014), their low signal-to-noise ratio and their short record. The reliance on hyperspectral instruments for SIF entails a compromise between spectral resolution, sounding footprint size, and spatial continuity. For OCO-2, SIF is retrieved as individual soundings with a relatively fine footprint of 1.3km×2.25km and large noise level (Frankenberg et al., 2014), but the wide gaps between neighboring swaths (~ 100km) imply that the data often has to be aggregated to a coarse spatial resolution (about 1° × 1°) at a monthly scale (Sun et al., 2018). While newer sensors such as TROPOMI provide near-continuous daily coverage of SIF observations thanks to a much wider swath and higher measurement frequency (Köhler et al., 2018), the wide range of solar zenith angle caused the large span of local solar time can potentially complicate the signal interpretation (Köhler et al., 2018). Ongoing efforts to generate a multidecadal SIF record by harmonizing data from overlapping sensors demonstrate the potential for back-calibrating earlier SIF retrievals with more accurate TROPOMI and OCO-2 targets. However, further understanding of the uncertainties introduced by different retrieval methods and calibration drifts of GOME-2 sensors is needed to address the remaining inconsistencies (Parazoo et al., 2019).

The spatiotemporal discontinuity, low spatial resolution and high noise of satellite SIF have motivated multiple studies to reconstruct global SIF maps using broadband reflectance and complementary information of land cover and surface meteorology (see Table S1 in the SI for a summary of SIF reconstruction studies). The overarching rationale for those reconstruction efforts is that satellite-based SIF contains information from $APAR_{chl}$, fluorescence yield $\Theta_F$, and escape ratio (see equation 3), with $APAR_{chl}$ being the dominant component of SIF variations (Zhang et al., 2018a). Both $APAR_{chl}$ (Knyazikhin et al., 1999; Roujean and Breon, 1995) and canopy radiative transfer (Zeng et al., 2019) can be partially captured by broadband reflectance (especially the red and near-infrared bands sensitive to canopy structure and greenness), creating a mechanistic linkage between SIF and reflectance channels. This assumption is further corroborated by the fact that SIF normalized by PAR is highly correlated with VIs (Zeng et al., 2022). As changes in fluorescence yield should mainly reflect short-term physiological signals, $\Theta_F$ can be theoretically ideally predicted with meteorological data. However, the effects of meteorological variables are insubstantial on the semi-monthly to monthly timescale as the SIF signal is mainly driven by slower-varying biochemical and canopy structural processes, making robust prediction of fluorescence yield challenging as $\Theta_F$ demonstrates minor variations under non-stressed conditions (Dechant et al., 2020; Zhang et al., 2018b). Furthermore, including meteorological data for predicting $\Theta_F$ may introduce circularity when linking SIF data with climate. A few studies found that including environmental variables as predictors did not significantly increase the prediction accuracy for SIF or $\Theta_F$ (Zhang et al., 2018a), which led the authors to construct models using reflectance data only (Gentine and Alemohammad, 2018; Wen et al., 2020; Zhang et al., 2018a). We also elected not to include land cover type as model predictors because previous study showed minimal performance gain from using land cover datasets in SIF prediction when reflectance measurements are already included (Li and Xiao, 2019).

The promise of reconstructing SIF using data-driven or machine learning algorithms has been confirmed by a recent intercomparison of four high-resolution global reconstructed SIF datasets (CSIF, GOSIF, LUE-SIF, and HSIF) with GPP measurements from eddy-covariance methods. The study found that all reconstructed SIF products are unequivocally better predictors of site-level GPP than remotely sensed vegetation indices such as NDVI and EVI (Shekhar et al., 2022). One remaining problem is the relatively short coverage of these reconstructed SIF products as they are trained with MODIS data, which are currently available for the last 20 years since the beginning of the MODIS era. Nevertheless, studies of phenological shifts, ecological droughts, and model calibration of biogeochemical processes would benefit from consistent, long-term, global observations of vegetation dynamics spanning multiple decades. In the absence of a long-term record of global SIF, researchers examining changes in ecological processes often turn to vegetation indices from earlier satellites to acquire a historical perspective of global change biology (Wang et al., 2020).

Advanced Very-High-Resolution Radiometer (AVHRR) has been the most common suite of Earth sensors used for long-term vegetation studies before the MODIS era, given its global coverage, daily repeat cycles, and, most importantly, continuous availability since 1982. While the reliance on AVHRR for vegetation monitoring has been gradually superseded by newer satellites offering superior resolution, accuracy and consistency, the 40 years of AVHRR record still represents an invaluable dataset for studying decadal vegetation trends particularly before 2000. There is therefore substantial potential to build a long-term reconstructed photosynthesis product by extending MODIS records to the 1980s using information from AVHRR. Nevertheless, some past studies have cautioned against a direct use of AVHRR data for long-term vegetation studies because it is plagued by several well-documented limitations, including spurious shifts in solar zenith angles (see Figure S1) due to orbital drifts (Doelling et al., 2016), inconsistencies between successive satellite sensors (Doelling et al., 2016; Frankenberg et al., 2021; Jiang et al., 2017), and the lack of on-board radiometric calibration (Doelling et al., 2016). These limitations necessitate carefully recalibrating the AVHRR reflectance.

In this study, we addressed the limitations in AVHRR reflectance by removing the orbital effects in AVHRR and cross-calibrating AVHRR against MODIS, in order to obtain a long-term consistent reflectance product. This new reflectance product was then used to define a Long-term Contiguous Solar-Induced Fluorescence (LCSIF) dataset from 1982-2022 at 0.05 degree spatial and biweekly temporal resolution. It is expected that the LCSIF product will be used as a new dataset for evaluating long-term global vegetation dynamics under a changing climate.

In the first part of the manuscript, we present the methodology. In the second part, we examine the calibration of AVHRR dataset and compared LCSIF with several established vegetation productivity proxies. We then discuss the long-term vegetation dynamics demonstrated by the reflectance and reconstructed SIF records, and highlight both the values and limitations of our product.

## 2 Methods

We calibrated and gap-filled AVHRR red and near-infrared reflectance channels from 1982-2022 to define a new temporally consistent Long-term Continuous REFlectance (LCREF-AVHRR) dataset at 0.05 degree spatial and biweekly temporal resolution. In addition, we also normalized, filtered, and gap-filled MODIS reflectance (LCREF-MODIS) to compare with concurrent AVHRR observations between 2001-2022. We trained a neural network for SIF reconstruction using only reflectance datasets and applied it separately on the AVHRR- and MODIS-based LCREF to reconstruct Long-term Contiguous Solar-Induced Fluorescence (LCSIF) from 1982-2022 (See Figure 1 for the schematics). Finally, we evaluated LCSIF against existing SIF products, VIs, and site-level GPP datasets.

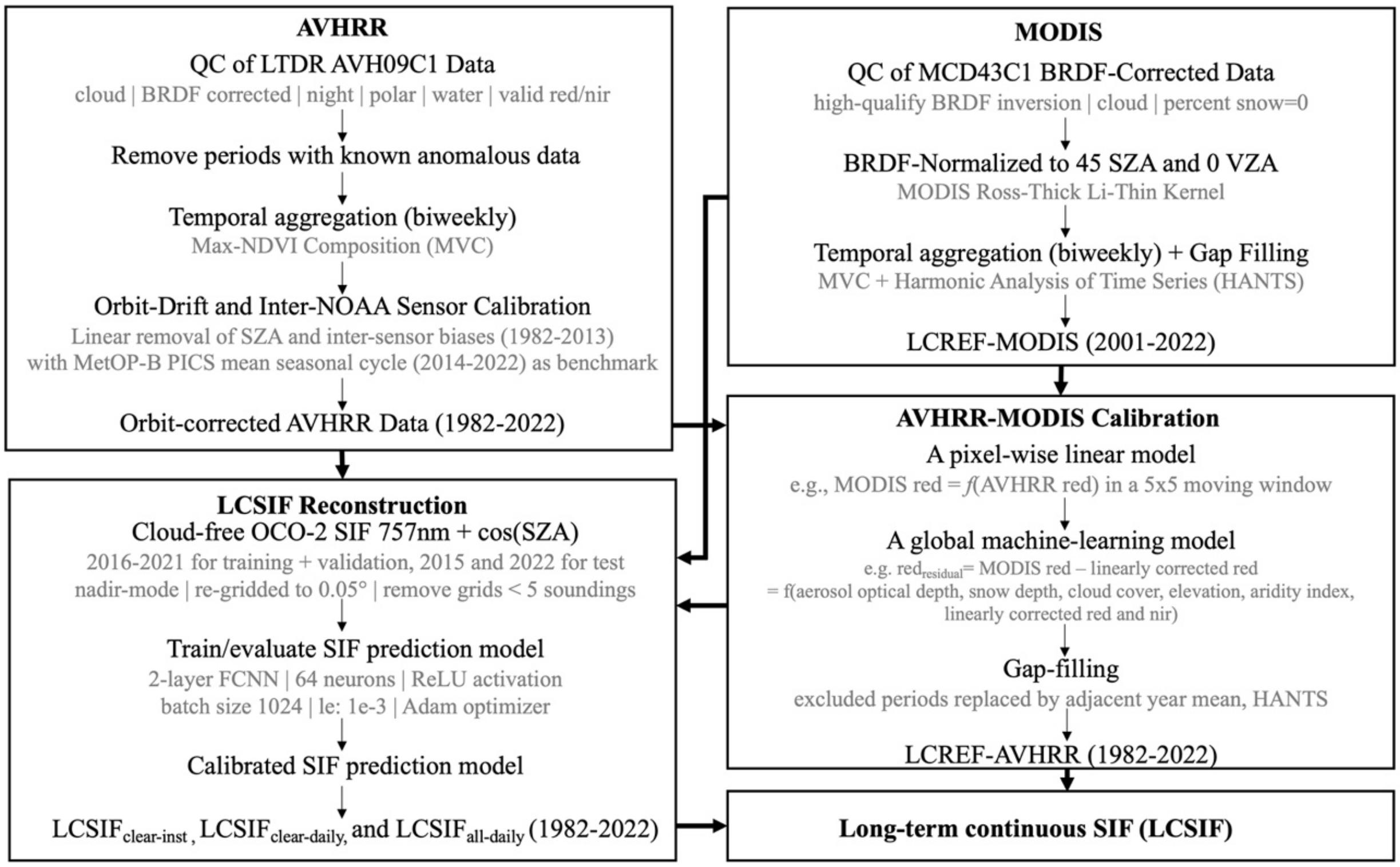

Figure 1: Flowchart showing the procedure of processing LCREF and the global reconstruction of LCSIF

*2.1 Datasets*

2.1.1 solar-induced chlorophyll fluorescence training data: OCO-2

The OCO-2 Solar Induced Chlorophyll Fluorescence Lite File V11r from 2015-2022 was used to train and test the machine learning model for the reconstruction of the LCSIF (Sun et al., 2018, 2017). We followed the preprocessing procedure used in the production of CSIF to prepare the SIF data for the machine learning model (Zhang et al., 2018a). The OCO-2 SIF retrievals were first filtered by measurement mode and the accompanying quality flags such that only the "best" quality clear-sky nadir mode measurements were used to develop the machine learning model. Next, the SIF data were aggregated to 0.05° daily grid cells to match the spatial resolution of the MODIS and AVHRR reflectance products (Zhang et al., 2018a). Aggregation also reduces the random noise in the SIF retrievals by a factor of 1/sqrt(n), where n is the number of soundings in each grid cell. To limit the random uncertainties in the training targets, we further excluded grid cells with fewer than five soundings (Zhang et al., 2018a). To examine whether the model performance was sensitive to the minimum OCO-2 soundings threshold, we also trained an alternative model using only grid cells with at least eight SIF retrievals. The SIF at 757nm was selected for the reconstruction of LCSIF because it has a stronger correlation with GPP than the SIF at 771nm (Li et al., 2018). We used the SIF data from 2016 to 2021 for model training (n=6,661,571 grid cells) and the data collected in the years 2015 and 2022 for testing (n=2,247,733 grid cells). We selected test data before and after the training period to avoid potential performance overestimation caused by temporal autocorrelation in the datasets. In addition, we randomly subsampled 20% of the training data to create a validation dataset for hyperparameter tuning during model development. Figure S2 shows the spatial distribution of the training and test samples, and Figure S3 shows the distribution of OCO-2 soundings per grid cell in the training dataset.

2.1.2 LTDR AVHRR surface reflectance

Multiple land-surface reflectance and VI products have been developed from AVHRR GAC level-1b, including the Pathfinder AVHRR Land (James and Kalluri, 1994), Global Inventory Modeling and Mapping Studies (GIMMS) (Tucker et al., 2005), the updated GIMMS3g dataset (Pinzon and Tucker, 2014), and the Land Long Term Data Record (LTDR) (Pedelty et al., 2007), each with a different data-sensor lineage and calibration procedure. For this study, we selected the LTDR product because it provides daily gridded reflectance measurements at the same spatial resolution as MCD43C1, and our preliminary analysis shows it suffers less from orbital effects than the GIMMS series. We downloaded the daily 0.05° LTDR V5 AVH09C1 dataset between 1982-2021 from the NASA LAADS DAAC archive (https://ladsweb.modaps.eosdis.nasa.gov/missions-and-measurements/applications/ltdr/#project-documentation) (Pedelty et al., 2007). The LTDR AVHRR surface product used the ocean and cloud vicarious radiometric calibration method to account for sensor degradations (Vermote and Kaufman, 1995), as well as the CLAVR-1 algorithm for cloud screening. Atmospheric corrections for molecular scattering, aerosol scattering, ozone absorption, and water-vapor absorption were applied in the data stream (Beck et al., 2011). To minimize the signal variations caused by changing sun-sensor geometry, LTDR AVHRR was normalized to a standard observation geometry (solar zenith angle at 45°, view zenith angle at 0°, and relative azimuthal angle at 0°) using the well-established Vermote-Justice-Bréon (VJB) method and the Bidirectional Reflectance Distribution Function (BRDF) parameters derived from MODIS data (Roger et al., 2021; Vermote et al., 2009; Villaescusa-Nadal et al., 2019). We filtered the LTDR pixels using the QC flags for cloud, snow, BRDF correction, night, polar, water, and valid Red/NIR surface reflectance to retain only high-quality day-time observations with good BRDF correction. Details about the AVHRR sensor origin for each period are listed in Table S2.

#### 2.1.3 MCD43C1 v061 MODIS/Terra+Aqua BRDF/Albedo Model Parameters

The MCD43C1 v061 MODIS/Terra+Aqua BRDF/Albedo Model Parameters were downloaded from LP DAAC. The MCD43C1 dataset was in a geographical coordinate grid with 0.05° spatial resolution based on a 16-day retrieval window. MCD43C1 contains weighting parameters used to drive the BRDF kernel for surface reflectance estimated from best-available observations over 16-day periods (Schaaf et al., 2002). We computed surface reflectance at the standard observation geometry (45° SZA, 0° VZA, and 0° RAA) using the MODIS Ross-thick Li-sparse-reciprocal BRDF kernel implemented in the SASKTRAN 1.8.2 package (Bourassa et al., 2008) to be consistent with the viewing geometry of the LTDR AVHRR dataset. We filtered the dataset to select cloud and snow-free pixels with high-quality BRDF-inversion flags (BRDF NBAR Quality ≤ 3).

#### 2.1.4 Ancillary dataset

We used the following datasets to account for the potential covariates that may affect the cross-calibration between MODIS and AVHRR reflectance values. Temperature, precipitation, and incoming solar radiation dataset were extracted from the CRUNCEP data (Viovy, 2018). Cloud cover and snow depth samples were retrieved from the ERA5-land reanalysis at 0.25° hourly dataset (Muñoz-Sabater et al., 2021). Aerosol optical depth (AOD) was obtained from the 3-hour 0.5 × 0.625 ° MERRA-2 reanalysis (Gelaro et al., 2017) to control for aerosol effects, the Global 30 Arc-Second Elevation dataset was used to control for the topography (GTOPO30) (Gesch et al., 1999), and the 4km Aridity Index (defined as potential precipitation divided by annual mean precipitation) computed from TerraClimate was used to control for surface aridity (as a measure for noises from background soils) (Abatzoglou et al., 2018). All datasets were aggregated to bi-weekly temporal resolution and resampled to 0.05° using a cubic function.

*2.2 Preprocessing of AVHRR and MODIS reflectance to remove inter-sensor and orbital effects*

We temporally aggregated the red and near-infrared bands of both MODIS and AVHRR to bi-weekly temporal resolution (1-15th for the first image, and 16th to the end of the month for the second image within the month) using the maximum value composite (MVC) based on NDVI (van Leeuwen et al., 1999). Three periods in the AVHRR record with either known instrumental malfunctions or anomalous observations were excluded from the calibration (See Figure S4 for the excluded periods, and SI Text 1 for a justification of anomalous data removal). As previous studies have confirmed MODIS to have good orbital stability and reliable radiometric calibration, we took it as ground truth to calibrate collocated AVHRR observations. While the LTDR surface reflectance product also applied the MODIS BRDF correction algorithm to rectify the spurious orbital drifts, past studies identified remaining orbital effects in the LTDR surface reflectance, which were propagated to down-stream NDVI, LAI, FPAR, and GPP products (Nagol et al., 2014; Tian et al., 2015). Furthermore, incomplete radiometric calibration results in systematic biases between individual AVHRR instruments (Latifovic et al., 2012), particularly between AVHRR-2 and AVHRR-3 sensors (See Figure S4). We extracted LTDR reflectance averaged over 10 Pseudo-invariant Calibration Sites (PICSs) to quantify the inter-sensor biases. PICSs are non-vegetated desert targets selected based on their radiometric stability and high reflectance values to enable the calibration of space-borne sensors (Bacour et al., 2019). Here, we assumed that the reflectance anomalies at PICSs mostly originated from sensor biases (Bacour et al., 2019; Frankenberg et al., 2021). Using the mean seasonal cycle of the most recent and radiometrically stable AVHRR-sensor onboard MetOP-B (MSC_MetOP-B_PICS$_{red\ or\ nir}$) as a reference, we computed the difference between the PICS-averaged reflectance series for all preceding AVHRR instruments with MSC_MetOP-B_PICS$_{red\ or\ nir}$ as a proxy of inter-sensor bias (Equation 5)

$$\delta_{red\ or\ nir} = PICS_{red\ or\ nir}\ (NOAA) - MSC_MetOP\text{-}B_PICS_{red\ or\ nir} \quad (5)$$

To remove the orbital effects and inter-sensor biases, we used an additional linear method to fit and then remove the contribution of solar zenith angle to the pixel-wise anomalies of surface reflectance in each of the red and near-infrared channels while retaining the explained anomalies by meteorological variables (Equation 6 and 7).

$$\rho_{red\ or\ nir} = a_1 \times SZA + a_2 \times \delta_{red\ or\ nir} + a_3 \times Ta + a_4 \times P + a_5 \times Rad + b \quad (6)$$

$$\rho'_{red\ or\ nir} = \rho_{red\ or\ nir} - a_1 \times SZA - a_2 \times \delta_{red\ or\ nir} \quad (7)$$

where $\rho_{red}$ and $\rho_{nir}$ are the anomalies in the LTDR Red and NIR channel, SZA is the anomaly of solar zenith angle at the time of acquisition, and $T_a$, P, and Rad are air temperature at 2 m, total precipitation, and incoming solar radiation from the CRUNCEP dataset. We fitted the equation for each of the 24 biweekly periods within a year using ordinary least squares regression (OLSR) and used the regression coefficients of SZA and $\delta_{red\ or\ nir}$ to remove their contributions to reflectance for each pixel, obtaining the SZA and sensor-bias corrected reflectance ($\rho'_{red\ or\ nir}$). These corrections were applied to AHVRR sensors onboard NOAA platforms between 1982-2013 only because the subsequent AVHRR sensor on MetOp-B (2014-2022) did not suffer from orbit drifts (Figure S4).

*2.3 Calibration of AVHRR reflectance bands against MODIS reflectance bands*

We leveraged both the local and global correlational structures between overlapping MODIS and AVHRR observations over 2000-2022 to learn the (nonlinear) functional relationship between the two sensors and their respective channels. We first considered the local temporal correlation between the two instruments using a 5 × 5 running window to fit a pixel-wise linear model that maps AVHRR reflectance to MODIS values. The implementation of running windows was made to ensure enough sample size, particularly for high-latitude regions affected by snow cover in cold seasons. The pixel-wise linear models were fitted and applied separately to the AVHRR sensors onboard NOAA satellites (1982-2013) and MetOp-B (2014-2022). Next, we computed the residual between the MODIS value and this linearly-corrected AVHRR value ($\gamma_{residual}$), which represents the remaining errors not captured by the locally adapted linear model. We assumed these differences could originate from the different non-linear responses of reflectance retrieved by MODIS and AVHRR instruments to environmental variates such as atmospheric (aerosol and cloud cover) and surface conditions (topography, snow and soil background noises).

$$\gamma_{residual} = \text{MODIS value} - \text{linearly-corrected AVHRR value} \quad (8)$$

Building on this assumption, we trained neural networks (See SI Text 2 for details) to capture the spatial relationship between $\gamma_{residual}$ and environmental covariates including aerosol optical depth (AOD), snow depth (SD), cloud cover (CC), elevation (ELE), climatological aridity index (AI), and the linearly corrected AVHRR red and NIR reflectance.

$$\gamma_{residual} = f(\text{AOD, SD, CC, ELE, AI, linear corrected AVHRR}_{\text{red and NIR}}) \quad (9)$$

We then applied both the pixel-wise linear model and the global-scale ML model for $\gamma_{residual}$ to AVHRR observations prior to the MODIS era so that the double calibration could be carried out over to the entire time series. Note that double-calibration was conducted respectively for each of the 24 biweekly periods within a year. After ML-correction, we filled each period excluded in section 2.2 with the seasonal mean of the preceding and following year. The remaining gaps in data were filled using the harmonic analysis of time series (HANTS) method, which applies a least squares curve fitting procedure for time series based on harmonic components of periodic functions (Roerink et al., 2003). We constructed two gap-filled reflectance products: LCREF-AVHRR using calibrated AVHRR reflectance from 1982-2022, and LCREF-MODIS using gap-filled and BRDF-normalized MODIS reflectance from 2001-2022. The two LCREF products were used for SIF reconstruction and the computation of various VIs for comparison in subsequent analyses.

*2.4 Reconstruction of LCSIF*

We then trained feedforward neural networks to map the red and near-infrared bands from BRDF-normalized daily MODIS reflectance to the aggregated daily OCO-2 SIF observations. As SIF correlates strongly with incoming solar radiation, we used cos(SZA) as an additional predictor for the proxy of radiation, where SZA was the average of solar zenith angles for each OCO-2 SIF sounding in a grid cell. All predictors (inputs of the neural network) were standardized to accelerate the model convergence. We experimented with learning rates of 0.001 and 0.0005, number of hidden layers between 1-3, and hidden layer dimensions of 8, 32, and 64. We trained the model for 40 epochs. While we also tried regularization techniques such as dropout and early stopping, we found minimal advantages of those regularizations given the absence of apparent overfitting when evaluating the model on the validation data (see section 2.1.1 for how the data was split between training, validation, and test). The hyperparameter set that performed best on the validation dataset was then used to retrain the model with both training and validation data combined to maximize the number of samples. The retrained model was tested on both the held-out test dataset as a whole and on test data from different land cover types. Finally, we used this model to generate a biweekly global SIF product at 0.05° spatial resolution for 1982-2021 using the calibrated LCREF data and the instantaneous solar zenith angle based on the predicted OCO-2 satellite overpass time (with a nominal equator crossing time set 1:36 pm) for different dates and latitudes. This product provides an estimate of the instantaneous clear-sky solar-induced fluorescence at various locations, comparable with the $SIF_{clear\text{-}inst}$ variable in the original CSIF product (Zhang et al., 2018a). In addition, we also produced a SIF product based on the mean solar zenith angle over each day as a proxy of daily mean clear-sky solar-induced fluorescence consistent with the $SIF_{clear\text{-}daily}$ variable in CSIF. The relation between $SIF_{clear\text{-}daily}$ and $SIF_{clear\text{-}inst}$ is based on a simple solar zenith angle correction with

$$SIF_{clear-daily} = SIF_{clear-inst} \times \frac{\cos(SZA)_{daily}}{\cos(SZA)_{inst}} = SIF_{clear-inst} \times \frac{\int_{t=t_0-12h}^{t=t_0+12h} |\cos(SZA(t)| dt}{\cos(SZA(t_0))} \quad (10)$$

in which $SZA(t_0)$ is the expected instantaneous SZA at the time of OCO-2 overpass. Furthermore, we produced a third product of daily-averaged SIF adjusted by ERA5 surface solar radiation downward (i.e., including cloud cover effects), defined as

$$SIF_{all-daily} = \frac{SIF_{clear-inst}}{PAR_{clear-inst}} \times PAR_{daily}^{ERA5} \quad (11)$$

where $PAR_{clear-inst}$ is an estimate of the clear-sky top-of-canopy radiation considering atmospheric scattering at the time of OCO-2 overpass (based on Appendix A1 in Zhang et al., 2018a), and $PAR_{daily}^{ERA5}$ is daily averaged surface solar radiation downward over each aggregation period of the reflectance dataset.

*2.5 Validation for LCREF and LCSIF*

We validated the orbital effects removal in the calibrated AVHRR reflectance record and the downstream LCSIF using 10 validation Pseudo-Invariant Calibration Sites (PICSs) in the Sahara Desert and on the Arabian Peninsula (see Table S3 for details). The validation PICSs differed from the 10 PICSs used for inter-sensor calibration in section 2.2 for an independent assessment. To understand how each calibration step affected the AVHRR reflectance record, we extracted reflectance values from the original, intermediate, and calibrated AVHRR variables. MODIS reflectance was also evaluated against the PICSs to confirm its temporal consistency. In addition, We examined whether PICS-level reflectance anomalies in the LCREF datasets were correlated with the global interannual variabilities in LCSIF because any spurious trends introduced by orbital drifts, if present, might be manifested at a global scale.

Next, we investigated the prediction accuracy of our machine learning model for mapping from reflectance to SIF values on the held-out test dataset. To investigate whether the model has robust predictive power for samples from different vegetation types, we divided the test dataset into different vegetation types based on the majority International Geosphere-Biosphere Programme (IGBP) land cover type in the MCD13C1 dataset from 2015-2022. To ensure that a sufficient number of samples are available for each vegetation type, we combined evergreen needleleaf forest (ENF) and deciduous needleleaf forest into a single "needleleaf forest" (NF) type, aggregated closed shrubland and open shrubland into a single "shrubland" (SH) type, and merged woody savanna (WSV) and savanna (SAV) into a combined "savanna" (SAV) type. We used the root-mean-square error (RMSE) and the coefficient of determination ($R^2$) between predicted and observed SIF as metrics for model performance.

We then evaluated the $R^2$ between LCSIF and site-level GPP measurements. We used the nighttime partition GPP estimate (GPP_NT_VUT_REF) at 165 eddy-covariance sites (See Table S4 for details) in the weekly-aggregated FLUXNET 2015 Tier One dataset as the benchmark, and we computed the $R^2$ between site-level GPP and predicted SIF at the overlapping pixel. To examine whether SIF could serve as a more robust proxy for GPP than other red and near-infrared bands-based VIs—NDVI (Tucker, 1979), kNDVI (Camps-Valls et al., 2021), and NIRv (Badgley et al., 2017)—we computed the $R^2$ between VIs (calculated from LCREF) and site-level GPP as a comparison. To account for the effects of sunlight variations on vegetation productivity not captured by reflectance measurements, we also compared LCSIF with NIRvP (computed as the product of NIRv and ERA-5 downward solar radiation). This recently developed proxy has been shown to robustly capture the spatiotemporal variations of SIF (Dechant et al., 2022).

Additionally, we examined the relationship between SIF and FLUXCOM, a machine-learning upscaling of site-level GPP. We selected the ANN-based RS+METEO FLUXCOM product, which used the climatology of various MODIS-derived datasets for spatial and seasonal variations and CRUNCEPv6 for temporal variabilities in meteorological forcings (Jung et al., 2020; Tramontana et al., 2016).

Our investigations encompassed both the original LTDR AVHRR dataset and LCREF-AVHRR to determine whether the calibration process enhances the correlations between satellite-derived and ground-based productivity estimates. We also conducted a comparative analysis of the correlation strength between LCREF-AVHRR-based variables and GPP before and after the year 2001. This allowed us to assess whether the benefits of the calibration procedure extended to the pre-MODIS era, utilizing a subset of 17 FLUXNET sites with at least three years of observations before and after the introduction of MODIS. All datasets were converted to a biweekly temporal resolution.

## 3 Results

### *3.1 Removal of AVHRR orbital effects*

Our analysis begins with examining the removal of AVHRR orbital effects at the 10 PICSs reserved for validation. As Figure 2 shows, the LTDR AVHRR reflectance over the calibration targets (shown in gray) exhibited large excursions punctuated by systematic discrepancies between AVHRR-2 and AVHRR-3 sensors, whereas the reflectance time series from the MODIS sensors (light orange) was much more stable. The orbital effects and inter-sensor bias correction (purple) notably enhanced the internal consistency within the AVHRR record. Moreover, implementing the pixel-wise linear and global machine-learning calibration with MODIS aligned the AVHRR record with its MODIS counterpart.

Our PICS analysis on the calibrated LCREF-AVHRR (Figure S5) also revealed that the magnitude of the residual orbital effects was small. The average annual detrended normalized anomalies for the calibrated red and near-infrared bands of the AVHRR reflectance were well within 1%. Furthermore, there were no significant correlations between the detrended normalized anomalies of reflectance values at the validation PICSs and the global anomalies of reconstructed LCSIF. Specifically, the correlations for red reflectance vs. LCSIF ($r=0.300$, $p=0.503$) and NIR reflectance vs. LCSIF ($r=0.300$, $p=0.095$) were found to be non-significant (as shown in Figure S5).

To examine whether the orbital effects removal was successfully applied at different locations, we also analyzed the pixel-wise correlations between the detrended anomalies of annual mean SZA and that of corrected reflectance bands and LCSIF. Results indicated that although the SZA excursions due to orbital drifts were more pronounced in the tropics than in high-latitude regions (Figure S6d), our calibration procedure removed spurious correlations due to SZA shifts in a spatially consistent manner. This was evident as the remaining correlations with SZA for LCREF and LCSIF were weak in most regions (Figure S6a-c).

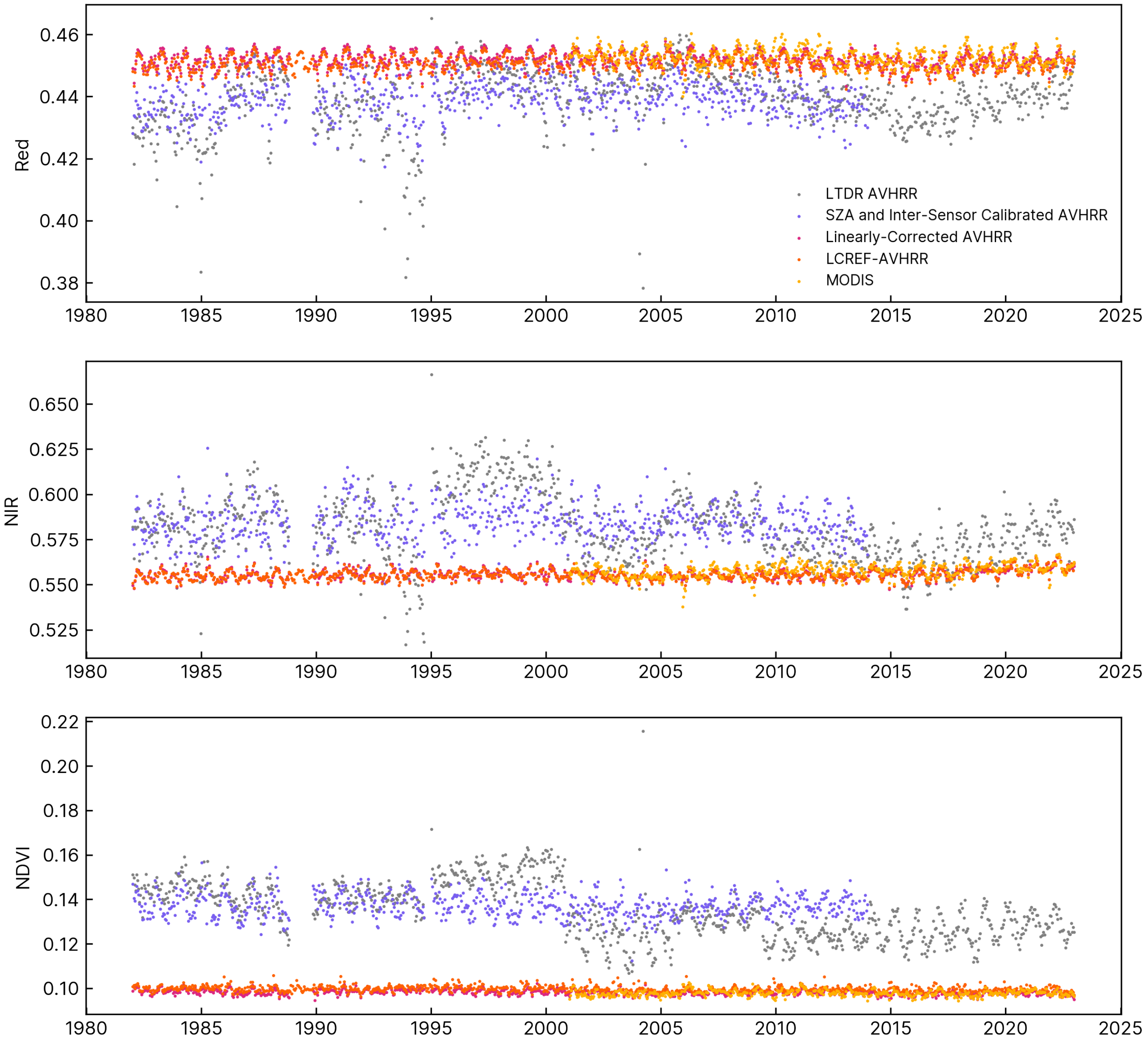

Figure 2: Comparison of MODIS and AVHRR reflectance against Pseudo-Invariant Calibration Sites (PICSs). In this figure, we present three panels depicting the median reflectance values observed at 10 validation PICSs for the red channel (top), the NIR channel (middle), and NDVI

(bottom). LCREF-AVHRR is contrasted with the original LTDR AVHRR, as well as intermediate products obtained after SZA + inter-sensor calibration and pixelwise linear-calibration with MODIS. Additionally, we include the median reflectance and NDVI series from MODIS at the PICSs for comparison.

*3.2 Model performance*

The machine learning model used for SIF reconstruction attained good prediction accuracy for estimating OCO-2 SIF, reaching a coefficient of determination ($R^2$) of 0.79 and RMSE of 0.19 mW $m^{-2}$ $nm^{-1}$ $sr^{-1}$ on the training dataset. The model also successfully captured the variations of OCO-2 SIF on the test dataset with an $R^2$ of 0.79 and RMSE of 0.19 mW $m^{-2}$ $nm^{-1}$ $sr^{-1}$, suggesting that the model showed minimal indication of overfitting (Figure S7). It is important to note that the predictive performance of our two-band model is comparable to that of the four-band algorithm used to produce CSIF ($R^2$=0.796 and RMSE=0.182 on the training dataset, $R^2$=0.786 and RMSE=0.177 on the test dataset) in terms of the correlation with OCO-2 soundings. We observed high correlations ($R^2$>0.65) between the observed and model-predicted SIF for deciduous broadleaf forest (DBF), mixed forest (MF), savanna (SAV), grassland (GRA), and cropland (CRO). However, we noticed a lower correlation ($R^2$=0.29) between predicted and observed SIF for the shrubland (SH), potentially due to the larger signal-to-noise ratios of SIF retrievals in sparely vegetated regions. In addition, we note slightly subpar correlations for needleleaf forest (NF, $R^2$=0.56) and evergreen broadleaf forest (EBF, $R^2$=0.43), probably caused by the weak seasonality in the canopy structure in these vegetation types. Sensitivity tests showed the model performance was robust to selecting different years for model evaluation (Figure S9), and the model performance showed minimal improvement when it was trained on grid cells with at least 8 OCO-2 soundings compared with the current approach.

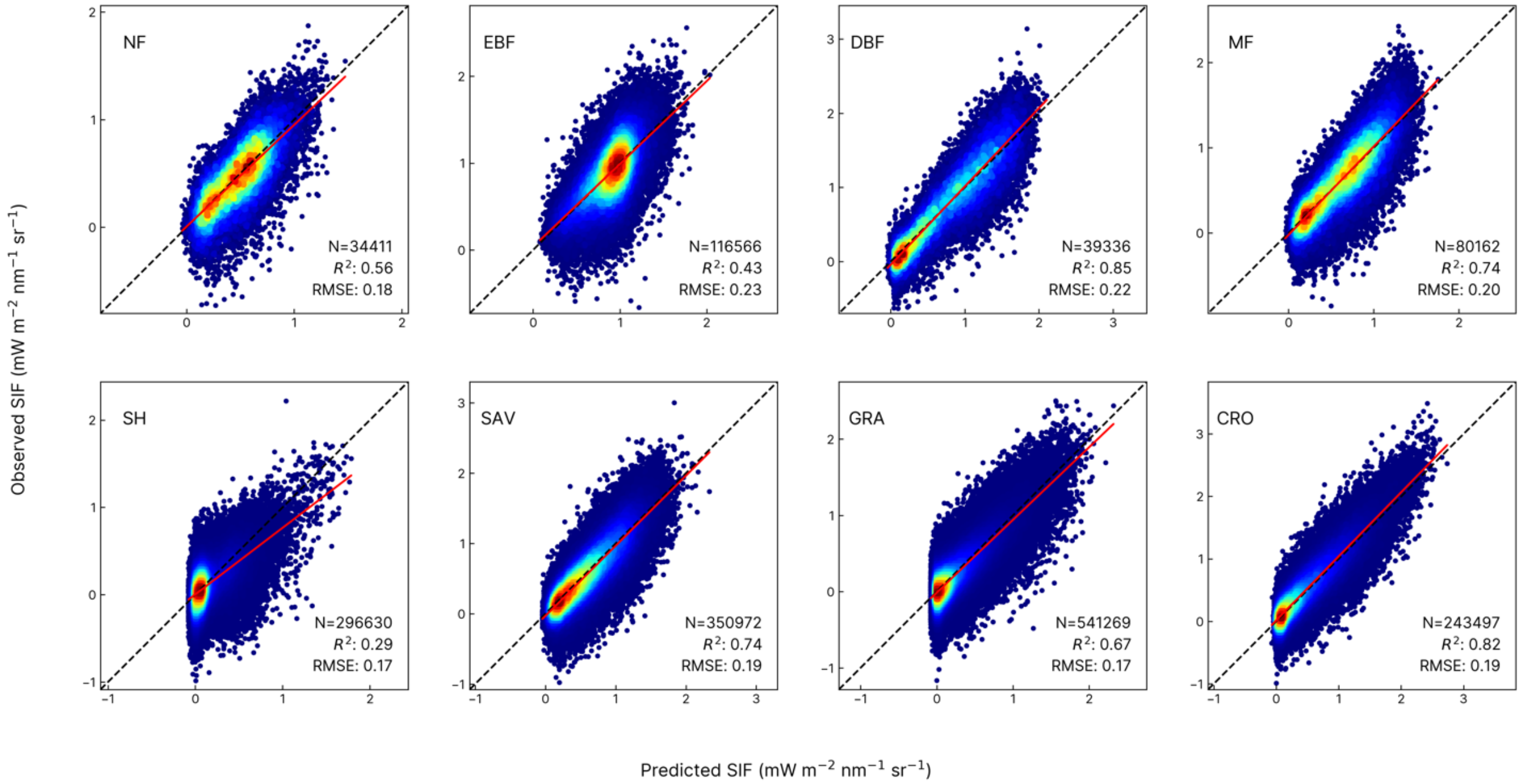

Figure 3: Evaluating the SIF prediction of LCSIF on the test dataset by IGBP land classification types. NF: Needleleaf Forest (evergreen and deciduous needleleaf forest); EBF: Evergreen Broadleaf Forest; DBF: Deciduous Broadleaf Forest; MF: Mixed Forest; SH: Shrubland (open and closed shrubland); SAV: Savanna (savanna and woody savanna); GRA: Grassland; CRO: Cropland (See Figure S8 for results on the training dataset). Each dot represents a daily grid cell of a particular land classification type, and the regression includes all grid cells of one specific land classification type over the years.

*3.3 Temporal and spatial pattern of LCSIF*

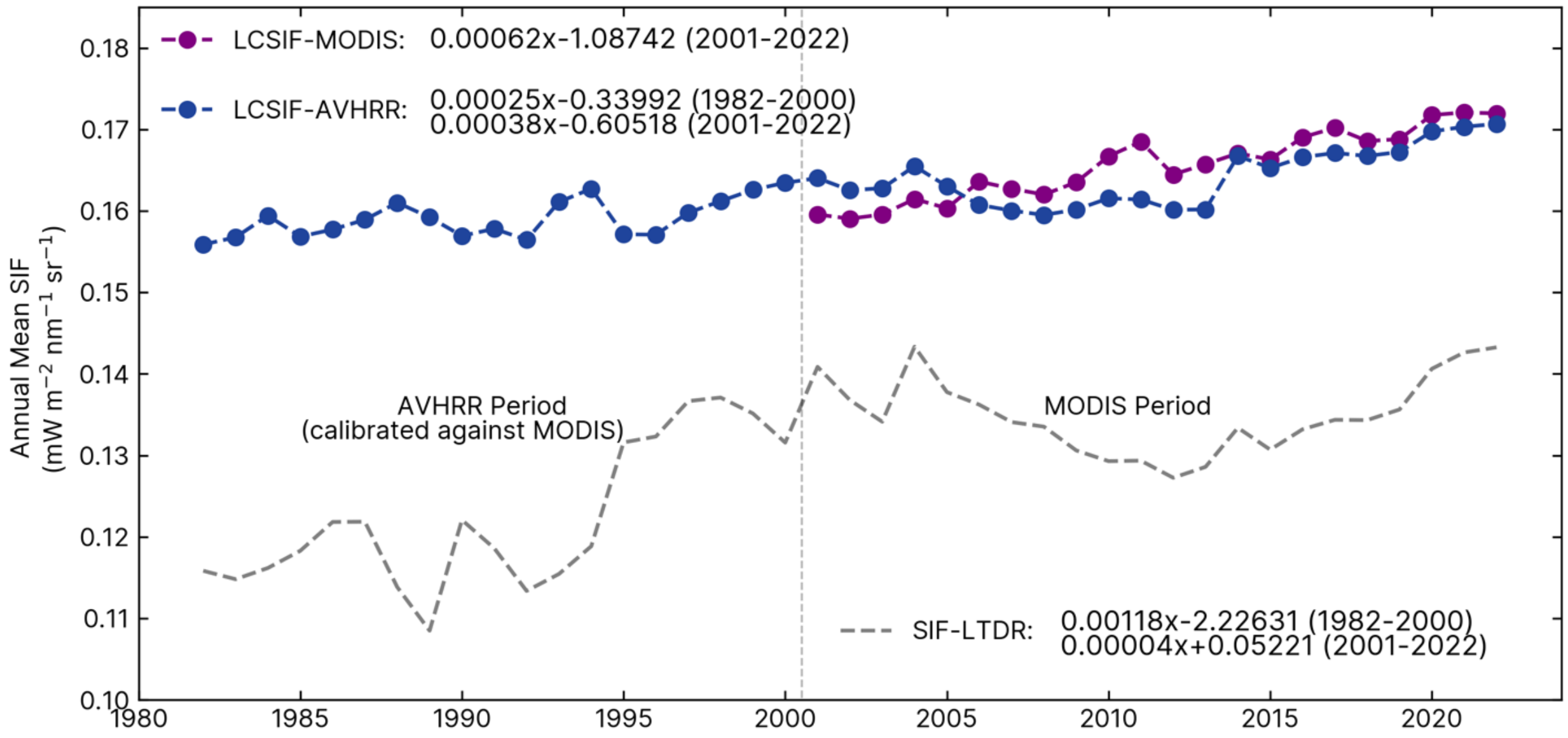

Figure 4: Annual mean growing season time series of AVHRR- and MODIS-based LCSIF. SIF directly reconstructed from the original LTDR dataset is also included for comparison. Pixels with SZA<0 (high-latitude winter) were assigned the value of 0 mW $m^{-2}$ $nm^{-1}$ $sr^{-1}$. The equations in the plot are linear least-square regressions of annual mean SIF for each dataset. We first computed the mean monthly SIF for each pixel before aggregating it to annual mean values. A common growing season data mask was applied to all datasets, and the pixels were area-weighted during spatial aggregation. We define the growing season as months where the climatology mean temperature is above 0 °C.

We first conducted a comparative assessment of the mean temporal trends in calibrated AVHRR and MODIS-based LCSIF reconstructions, as depicted in Figure 4. LCSIF-AVHRR displayed a global increasing trend of 0.0025 mW $m^{-2}$ $nm^{-1}$ $sr^{-1}$ per decade during the period from 1982 to 2000. Between 2001 and 2022, the trend in LCSIF-AVHRR appeared to dip slightly in the mid-2000s before resuming an upward trajectory in the 2010s, resulting in an overall increasing rate of 0.0038 mW $m^{-2}$ $nm^{-1}$ $sr^{-1}$ per decade. In contrast, LCSIF-MODIS exhibited a more rapid upward trend, with a rate of 0.0062 mW $m^{-2}$ $nm^{-1}$ $sr^{-1}$ per decade.

As a reference point for comparison, we examined SIF directly reconstructed using LTDR AVHRR data. SIF-LTDR experienced a substantial increase between 1982 and 2000, with a rate of 0.0118 mW $m^{-2}$ $nm^{-1}$ $sr^{-1}$ per decade, accompanied by notable fluctuations. However, the overall trend of SIF-LTDR was nearly flat from 2001 to 2022, driven by a significant decline in the 2000s, followed by a recovery in the subsequent decade. When compared with SIF-LTDR, SIF reconstructed using calibrated AVHRR reflectance exhibited considerably smaller interannual variabilities and more closely aligned with MODIS-derived values in both absolute value and trend.

To assess the impact of long-term solar radiation shifts (Wild et al., 2009), we compared the normalized anomalies of clear-sky daily ($LCSIF_{clear\text{-}daily}$) and all-sky daily SIF ($LCSIF_{all\text{-}daily}$). Although the mean value of all-sky SIF is only ~75% of the clear-sky SIF value, their normalized anomalies are highly correlated ($r>0.98$), suggesting that the trends in surface incoming solar radiation has only a minor impact on the long-term SIF trends.

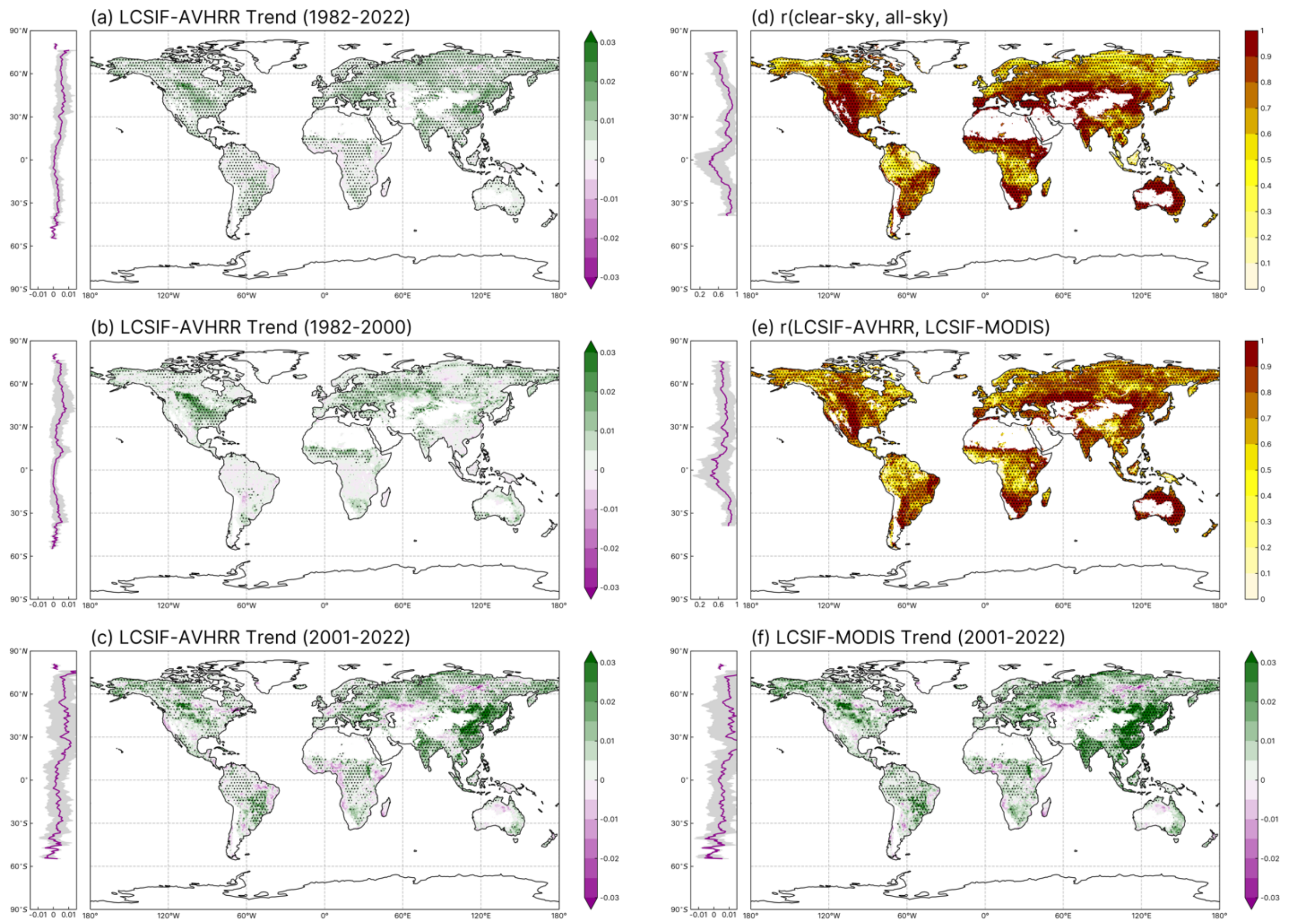

Figure 5: Spatial trends and correlations in AVHRR and MODIS-based LCSIF. (a) Depicts the spatial pattern of AVHRR-based LCSIF trends for the entire record spanning from 1982 to 2022, while panels (b) and (c) separate the trends for the periods 1982-2000 and 2001-2022, respectively. (d) Illustrates the annual-level correlation between clear-sky and all-sky LCSIF, and (e) provides the correlation between clear-sky and all-sky LCSIF-AVHRR. (f) Presents the MODIS-based LCSIF trends for the period 2001-2022. Each subplot includes an inset on the left, displaying the latitudinal distribution of trends or correlations along with their associated standard deviations. Areas with a multi-year average SIF < 0.03 mW $m^{-2}$ $sr^{-1}$ $nm^{-1}$ have been excluded and are represented in white. Significance levels for both trends and correlations are determined through pixel-wise least square regression, with alpha set at 0.05. All trends and correlations are computed from annual growing season mean.

We next examined the spatial pattern of the LCSIF trends for both AVHRR and MODIS-based versions. During the 1982-2000 period, an increasing LCSIF-AVHRR trend was predominantly observed in eastern Europe, the Sahel, southern Africa, as well as in the Midwest and southeastern regions of the United States (Figure 5b). While the trend of enhanced productivity in most of these regions continued into the early 21$^{st}$ century, the most notable LCSIF increase during the 2001-2022 period was found in eastern China, India, and the soybean belt of South America (Figure 5c). These regions with amplified SIF trends were the primary contributors to the accelerated global trend from 1982-2000 to 2001-2022 (Figure 4). Furthermore, a rise in LCSIF was also evident across much of the boreal forest during the 2001-2022 period. By contrast, a slight browning pattern was noted in the eastern Amazon (Figure 5c, f).

LCSIF-AVHRR and LCSIF-MODIS showed a generally consistent spatial trend during their overlapping period (Figure 5c, f), despite the mild decline of LCSIF-AVHRR in mid-2000s, as observed in Figure 4. The interannual variabilities of MODIS and AVHRR-based SIF reconstructions were positively correlated over much of the global vegetated surface. However, the strength of the correlation appeared to be lower in the tropics and parts of the northern forest (Figure 5e). Accounting for the cloud cover and aerosol effects on surface solar radiation led to a moderate reduction in the clear-sky vs. all-sky LCSIF-AVHRR correlations in the wet tropics and the high-latitude boreal region (Figure 5d).

To ensure the robustness of the results, we also computed the pixel-wise SIF trend using the Theil-Sen slope estimator and tested the significance of the slope with the Hamed & Rao modified Mann-Kendall (MK) trend test, a robust non-parametric test to detect the monotonic upward and downward trend in serial data. The results from the MK test (Figure S10) were consistent with the spatial patterns in Figure 5, confirming the statistical robustness of the derived trends.

### *3.4 Comparison of LCSIF with site-level GPP*

We evaluated the $R^2$ between LCSIF and GPP estimated from eddy-covariance sites in the FLUXNET2015 datasets to investigate whether the LCSIF is a reliable proxy for GPP. As a benchmark for comparison, we computed the $R^2$ between site-based GPP and satellite-based LCSIF and of three widely-adopted red and NIR band-based VIs: Normalized Difference Vegetation Index (NDVI) (Tucker, 1979), Near-infrared Reflectance of vegetation (NIRv) (Badgley et al., 2017), and kernel NDVI (kNDVI) (Camps-Valls et al., 2021). Additionally, we also included a comparison of LCSIF with the recently-developed NIRvP, which is computed as the product between NIRv and ERA-5 incoming solar radiation.

Our results demonstrated that LCSIF outperformed VIs that solely relied on reflectance information (NDVI, NIRv, kNDVI) across a wide range of vegetated land cover types. However, the capability of reconstructed SIF in capturing GPP variations was on par with NIRvP, which benefits from the additional information about incoming solar radiation (Figure 6). The reconstructed SIF products exhibited their greatest advantage over the VIs in the evergreen broadleaf forest (EBF) and needleleaf forest (NF), likely due to these vegetation classes having small seasonal canopy structure variations, which are the primary signals of VI variabilities. Remarkably, we did not observe a significant improvement in $R^2$ after adjusting clear-sky daily SIF with all-sky incoming radiation, suggesting that cloud cover variations had only have a minor impact on a biweekly timescale. A further comparison with FLUXCOM GPP suggested that LCSIF-AVHRR had generally comparable correlation with site-level GPP as the state-of-the-art FLUXCOM upscaling product, despite being a much more parsimonious product using of only two reflectance bands from AVHRR as inputs.

We also found the calibration procedure significantly improved the correlation of the satellite-derived productivity proxies with site-level GPP at MF, NF, and SAV sites (Figure 6). Nevertheless, the MODIS-based VIs and SIF reconstructions still demonstrated more robust correlation with site-level GPP than the calibrated AVHRR-based products at some EBF, GRA, and SH sites (Figure S11), although the magnitude of the differences were marginal in many cases.

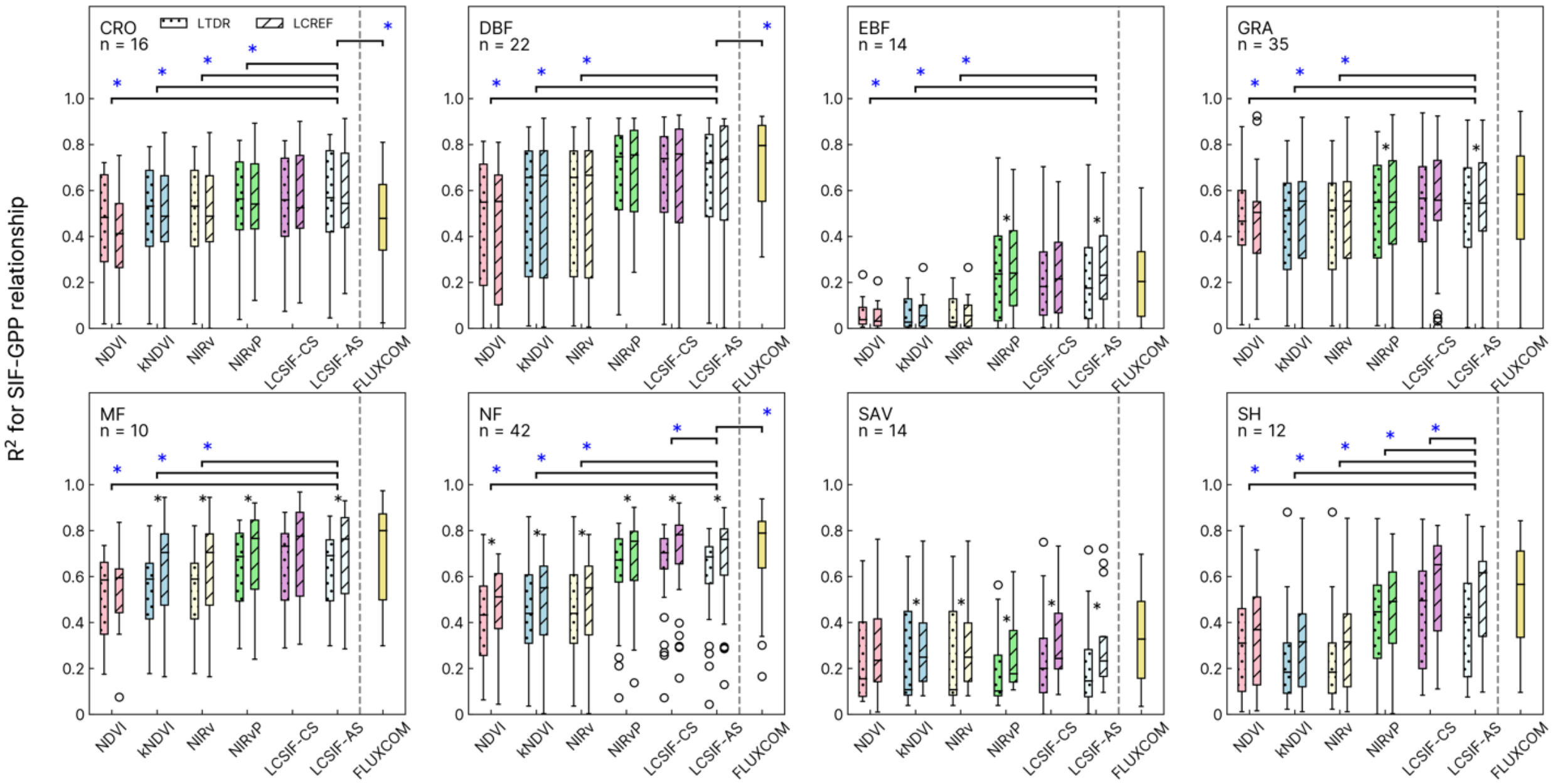

Figure 6: Comparative analysis of the calibrated AVHRR and the original LTDR AVHRR based VI-GPP and SIF-GPP correlations at eddy-covariance sites. For this analysis, we used nighttime partitioned GPP in the FLUXNET 2015 dataset starting from 2001. A blue asterisk signifies whether the correlation with GPP is stronger in the case of all-sky LCSIF (LCSIF-AS) compared to a reflectance-only VI, as determined by a paired t-test with a one-tailed p-value <0.05. It also signifies whether there is a difference in the correlation with FLUXNET GPP between all-sky

LCSIF (LCSIF-AS), FLUXCOM GPP, NIRvP, or clear-sky LCSIF (LCSIF-CS), as determined by a paired t-test with a two-tailed $p<0.05$. The black asterisk indicates, for each type of VI or SIF variable, whether the calibrated LCREF-based version has a higher correlation with site-level GPP compared with the LTDR AVHRR-based product (one-tailed $p<0.05$).

To examine whether applying the calibration derived from the MODIS-era to AVHRR observations before 2001 might lead to a degradation in the reliability of LCSIF, we subsampled 17 FLUXNET (encompassing three land cover classes) sites with at least three years of measurements both before and since 2001, and compared whether there was a systematic difference in $R^2$ between the two periods (Figure 7). We found, in general, no significant difference in correlation between the satellite-derived and ground-based productivity observations in the periods before and since 2001, except for NDVI and NIRv at evergreen needleleaf forest (ENF) sites. Nevertheless, we caution that the number of available site-year samples during the AVHRR-only period was less than half of the MODIS period, and the differences in $R^2$ could also be caused by selection biases or a lack of statistical convergence. An additional comparison with VIs indicated that, even within the AVHRR period, LCSIF has a greater or consistent correlation with GPP than VIs do (paired t-test, one-tailed $p<0.05$ in all cases). These results suggested that although there might be remaining errors in the calibrated AVHRR reflectance dataset, our SIF reconstruction algorithm was able to generate a “best possible” estimate of GPP given the data limitations.

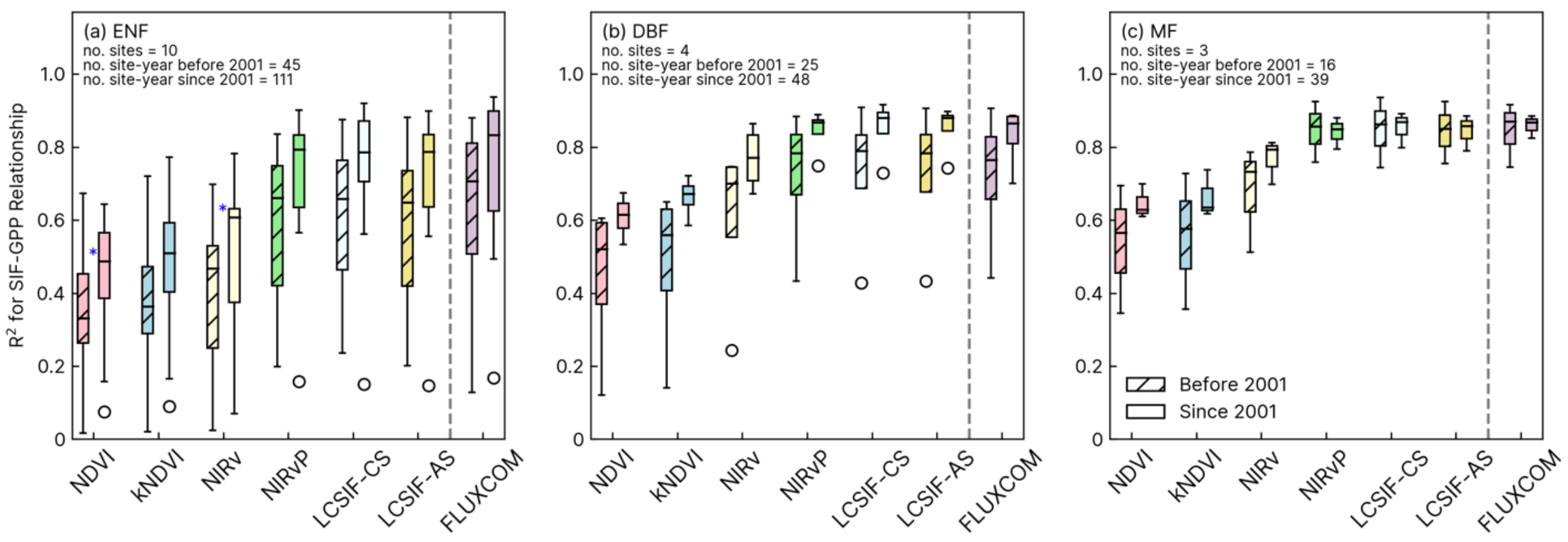

Figure 7 a-c: Comparing the correlation between LCREF-AVHRR derived productivity indices with site-level GPP before and since 2001. For this comparison, we selected 17 sites with at least 3 years of observations before and after the introduction of MODIS. The blue asterisk in this plot signifies whether the correlation with site-level GPP is significantly higher since the MODIS era than the period before (one-tailed $p<0.05$).

## 4 Discussion

In this study, we reconstructed a long-term biweekly global SIF dataset at 0.05° resolution by cross-calibrating the AVHRR and MODIS reflectance and extending the MODIS records to the pre-2000 period. We showed that the two-band model for SIF reconstruction demonstrated a performance comparable with the four-band original algorithm used in CSIF, yet with the advantage of being extendable to the AVHRR period. The calibrated AVHRR-based reconstruction of LCSIF showed generally consistent spatiotemporal patterns when compared with the MODIS-based reconstruction, and its capability to describe plant productivity dynamics was demonstrated with the high correlation with site-level GPP estimates compared to reflectance-based VIs (NDVI, NIRv, kNDVI). These results suggest that LCSIF can be a promising indicator of vegetation productivity, but users are also given the choice to compute any red and near-infrared reflectance-based vegetation index suitable for their analysis using the calibrated reflectance dataset.

Our effort to calibrate the AVHRR reflectance against the MODIS record also addresses some of the well-known limitations in this earlier remote sensing product. Unlike newer instruments with on-board solar diffusers to correct for the degradation in the visible and near-infrared sensors, AVHRR was not equipped with an on-board calibration instrument, making it more susceptible to fluctuations in the solar channel signals (Doelling et al., 2016). Furthermore, the overlap of the broad near-infrared channel of AVHRR with the absorption bands of $O_3$ and $H_2O$ resulted in more significant atmospheric contaminations (Tanre et al., 1992), which complicated the comparisons with reflectance measurements from narrower channel instruments. These challenges were partially addressed in our paper by using a double-correction procedure that leverages both the temporal and spatial correlations between AVHRR and MODIS observations. The local linear regression captured the temporal relationship between the two instruments, whereas the global neural network fitted to the regression residuals helped us alleviate the spatial errors introduced by atmospheric and surface covariates.

One of the most chronic issues in the history of AVHRR calibration is the lack of a propulsion system on earlier NOAA satellites, so that the AVHRR instruments could not maintain a consistent nominal equatorial crossing time, and each of these satellites eventually drifted from a sun-synchronous orbit to a terminator orbit over the lifespan of a few years (Doelling et al., 2016; Frankenberg et al., 2021). The orbital drift led to a spurious shift in the solar zenith angle at the time of acquisition within each AVHRR sensor that interferes with the reflectance signal. As the AVHRR record was derived from multiple NOAA satellites, each with its own orbital decay and calibration parameters, both the solar zenith angle at the time of acquisition and the reflectance records showed abrupt changes when the satellite source was switched from one satellite to the next (Frankenberg et al., 2021). This orbital drift issue in AVHRR underlies much of the recent debates as to whether there has been a greening or browning trend in North American boreal forest (Alcaraz-Segura et al., 2010), and whether there has been a recent global decline of $CO_2$ fertilization effect on vegetation photosynthesis (Frankenberg et al., 2021; Wang et al., 2020). While indices such as NDVI can partially mitigate the BRDF effects by canceling the residual BRDF effects through a normalized difference (Kaufmann et al., 2000), contaminations by solar zenith angle trends are often pronounced in indices such as NIRv that cannot alleviate the orbital effects by taking a ratio between reflectance channels (Badgley et al., 2017; Frankenberg et al., 2021; Zeng et al., 2022). Given that SIF and NIRv are highly correlated ($r \sim 0.86$), we expect that the orbital effects in the AVHRR reflectance records will likely propagate to the reconstructed SIF unless the input signals are properly harmonized (Zeng et al., 2022). Evaluation at PICSs sites suggest that our method successfully removed most of the spurious signal caused by orbital shifts in reflectance channels of AVHRR, and the resulting datasets showed no abrupt shifts between successive AVHRR sensors. We note that those reflectance datasets are made available to the public.

This temporally consistent and spatially contiguous SIF product points to an increasing trend of terrestrial ecosystem productivity over the past 40 years. LCSIF shows different regional patterns of productivity trends during the pre-2001 and post-2001 periods. The pre-2001 period is characterized by a productivity increase in the boreal forest, the Sahel, and the eastern US. These patterns are consistent with previous studies that associate regional greening with the increase in growing season and peak season vegetation activity in the boreal forest when soil moisture is not limiting, in Sahel with the increased precipitation between the 1980s and 1990s (Olsson et al., 2005), and in the Eastern US associated with natural forest regrowth (Ramankutty et al., 2010; Zhu et al., 2016). Noticeably, our product also shows a significant trend of greening/enhanced productivity in highly developed agricultural regions such as the US Midwest. Previous studies have found that intense modern management (e.g., irrigation, fertilizer use) has created a substantial increase in crop yield in US Midwest since the 1980s (Gao et al., 2019). Further, the significant positive trend in the LCSIF dataset seems to suggest that LCSIF is less susceptible to saturation in highly developed agricultural regions than greenness-based indices.

During the post-2001 period, the intense SIF increase in eastern China and India has been associated with large-scale reforestation and agricultural intensification, respectively, as indicated by a recent LAI-based study (Chen et al., 2019). The land use land cover change (LULCC)-induced regional increase of SIF adds to that caused by climate change and rising atmospheric $CO_2$, likely underlying the accelerated enhancement of ecosystem productivity prior and after 2000. It should be noted that LCSIF also shows regional SIF reduction in the eastern Amazon since 2001 that could be associated with local deforestation and cropland expansion. We find that the total area that shows a trend of enhanced productivity is much greater when examining the entire 1982-2022 record than when examining the MODIS and AVHRR periods individually. These results are consistent with a recent study that used robust statistical methods to control for spatial and temporal autocorrelations when detecting greening and browning patterns in satellite-derived vegetation records (Cortés et al., 2021).

Despite the general agreements between the calibrated AVHRR and MODIS-based SIF reconstruction during 2001-2022, we recognized a period with divergent trends in the mid-2000s during which LCSIF-MODIS demonstrated a consistent increase, whereas LCSIF-AVHRR exhibited a slight decline (Figure 4). While our calibration efforts vastly reduced the magnitude of the differences between the AVHRR and MODIS products compared with the original LTDR AVHRR, we found this difference persisted in the calibrated product, leading to a more subdued increasing trend of the AVHRR-based LCSIF between 2001-2022. We note that applying stringent snow masks, focusing on growing season pixels, and considering non-linear seasonal biases between the reflectance channels could not eliminate the discrepancy. We also believe the solar zenith drift issue to be an insufficient explanation for the difference because the N16, N18, and N19 satellites of this period demonstrated a much smaller range of SZA drifts than the earlier NOAA platforms (see Figure S1 & S4).

In a recent effort to generate a spatiotemporally consistent global dataset of GIMMS NDVI (PKU GIMMS NDVI3g) from 1982-2022, the authors also identified incongruous trends between MODIS and AVHRR during the mid-2000s despite extensive calibration with Landsat (Li et al., 2023). This latest GIMMS NDVI dataset tackled the issue by blending the calibrated GIMMS NDVI3g product derived from AVHRR between 1982 and 2002 and MODIS NDVI product between 2003 and 2022 to construct the full NDVI time series. Considering the overall superior quality of the MODIS dataset compared with AVHRR and its higher correlation with site-level GPP (Figure S11), we believe the consolidation decision is a justifiable compromise. Users of the LCSIF and LCREF datasets will also have the option to choose between MODIS or calibrated AVHRR-based reconstruction starting from 2001. Nevertheless, we acknowledge that the incompatible trend issue reveals the limitations of statistical calibration approaches, and further investigation is needed to understand the mechanistic causes of this discrepancy.

Overall, we believe the long-term reflectance and reconstructed SIF datasets generated from this study can provide a valuable addition to the data assets for studying long-term vegetation dynamics. We propose that the potential applications of the LCSIF and LCREF datasets include 1) reevaluating the regional greening vs. browning trend under climate change given the improved temporal consistency of the dataset (Zhu et al., 2016); 2) tracking phenology changes and detecting differences in the seasonality of productivity vs. canopy greenness; 3) constraining global biogeochemical models through data assimilation (Quetin et al., 2023); 4) establishing a baseline for detecting vegetation stress to be combined with meteorological and observed SIF values during droughts and heatwaves (Zhang et al., 2018a); 5) assessing vegetation resilience after natural or anthropogenic disturbance effects including fire and deforestation.

Compared with traditional VIs that mainly capture changes in vegetation canopy structure, SIF is often considered a more direct proxy of the physiological activity of photosynthesis. Indeed, we find that our LCSIF product is more highly correlated with site-level GPP estimates than other VIs, and it can sometimes outperform machine-learning upscaling despite being a much more parsimonious model. Nevertheless, as the model only uses reflectance datasets as predictors, we caution against directly interpreting the patterns in LCSIF and other reconstructed SIF datasets as a physiological signal and attributing their variations to physiological stressors. In particular, we do not recommend using LCSIF to retrieve the fluorescence quantum yield $\Phi_F$ because the machine learning contains no independent constraint for quantum yield. Instead, the reconstructed SIF should be considered as a proxy of vegetation activity and especially of chlorophyll available photosynthetically active radiation. The high correlation between OCO-2 SIF and GPP allows the reconstructed SIF product to retain more information about vegetation activity than VIs computed with a fixed formula such as NDVI, NIRv, and kNDVI, but the trends and variability in the reconstructed SIF ultimately depend on the quality of underlying reflectance datasets. The current dataset availability limits the spatial resolution of the dataset at 0.05° resolution, which means potential users should be cautious when applying the dataset in ecosystems with high spatial heterogeneities. This intermediate product will be useful for historical vegetation trend analysis while we await new sensors to provide temporally and spatially continuous SIF observations from space.

## 5 Conclusion

This study aims to provide a temporally consistent global vegetation productivity dataset by using calibrated AVHRR and MODIS reflectance between 1982-2022 to reconstruct OCO-2 solar-induced chlorophyll fluorescence. Both the AVHRR and MODIS-based LCSIF reconstructions indicate an overall accelerating increase in global SIF trend in the past four decades, despite a period of trend disagreements in the mid-2000s. Validations of the calibrated reflectance and reconstructed SIF datasets demonstrate the successful removal of orbital effects, the improved harmonization of reflectance channels between MODIS and AVHRR, and promising correlation with eddy-covariance based GPP estimates across diverse vegetation types. We envision the LCSIF and the LCREF datasets will be useful for identifying hotspots in vegetation greening, tracking phenological shifts in vegetation productivity and canopy structure, monitoring vegetation resistance and resilience against disturbance events (e.g., drought and fire), and informing the strength of terrestrial carbon sinks via assimilation into terrestrial biosphere models.

## Description of Author's Responsibilities

P.G., J.F., X.L. conceptualized the project. C.J and Y.R. developed the method for the detection and removal of orbital effects in AVHRR. X.L. and J.F. carried out the LCREF calibration. J.F. trained the SIF model and generated the LCSIF products. J.F. and X.L evaluated the results. J.F. wrote the initial manuscript. P.G. secured funding for the project. All authors are involved in discussing the results and editing the manuscript.

## Funding

Jianing Fang, Xu Lian, Youngryel Ryu, Sungchan Jeong, and Pierre Gentine acknowledge support from the LEMONTREE (Land Ecosystem Models based on New Theory, obseRvations and ExperimEnts) project, funded through the generosity of Eric and Wendy Schmidt by recommendation of the Schmidt Futures programme. Jianing Fang, Xu Lian, and Pierre Gentine also acknowledge funding from the European Research Council grant USMILE (ERC CU18-3746), and National Science Foundation Science and Technology Center LEAP, Learning the Earth with Artificial intelligence and Physics (AGS-2019625).

## Acknowledgement

We would also like to thank Weiwei Zhan and Yu Huang for sharing their initial results of reconstructing SIF using AVHRR.

## Code & Data Availability

The main data output of this study is the Long-term Contiguous Solar-Induced Fluorescence reconstructed using calibrated AVHRR record (LCSIF-AVHRR) from 1982-2022. We provide reconstruction of instantaneous clear-sky SIF ($SIF_{clear-inst}$), daily mean clear-sky SIF ($SIF_{all-daily}$), and daily-mean all-sky SIF ($SIF_{all-daily}$) as three proxies of vegetation activity.

The latest version of LCSIF (v3.1) can be accessed at https://doi.org/10.5281/zenodo.11654468 for 1982-2000, and https://doi.org/10.5281/zenodo.11906676 for 2001-2022. In addition, we also provide the calibrated AVHRR ned and NIR reflectance used to generate LCSIF-AVHRR as long-term continuous reflectance record (LCREF-AVHRR), publicly available at https://doi.org/10.5281/zenodo.11905960. Prospective users can readily compute common red and NIR based vegetation indices such NDVI, kNDVI, and NIRv suitable for their applications using LCREF-AVHRR. We also made available MODIS-based LCSIF and LCREF to facilitate comparison with AVHRR for the overlapping period from 2001-2022, with the two LCSIF product generated using the same model. The file structures of LCSIF-MODIS (https://doi.org/10.5281/zenodo.11658089) and LCREF-MODIS (https://doi.org/10.5281/zenodo.11657459) are identical to the AVHRR versions. The reflectance datasets are normalized to the same sun-object-sensor geometry as the AVHRR product and gap-filled using the same approach. All dataset outputs from this study are available at 0.05° spatial resolution and biweekly temporal resolution in NetCDF format. Each month is divided into two files, with the first file "a" representative of the 1st day to the 15th day of a month, and the second file "b" representative of the 16th day to the last day of a month. The code used to generate and evaluate the LCSIF data is available at https://github.com/JianingFang/longterm_continous_sif.

**List of Figure Captions**

**Supplementary Material for Reconstruction of a Long-term spatially Contiguous Solar-Induced Fluorescence (LCSIF) over 1982-2022**

Jianing Fang, Xu Lian, Youngryel Ryu, Sungchan Jeong, Chongya Jiang, Pierre Gentine

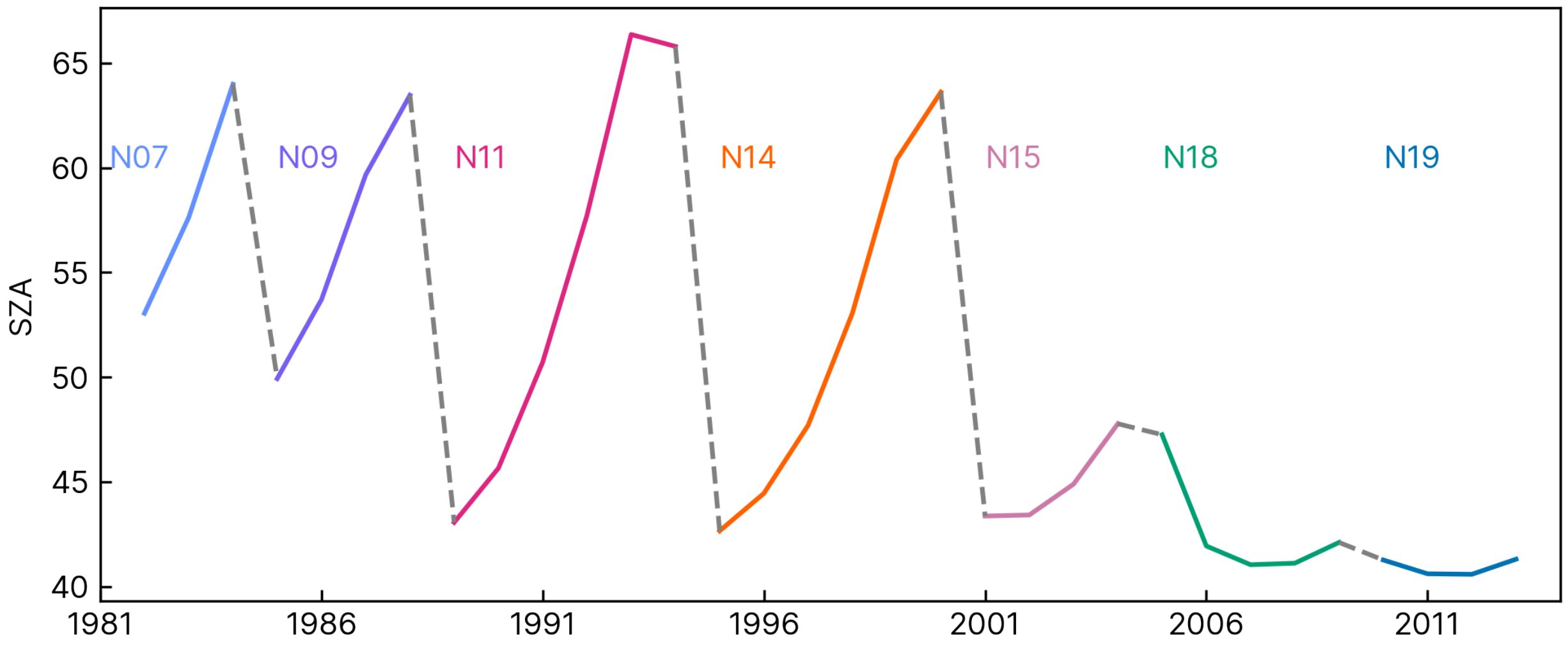

Figure S1: Annual mean solar zenith angle at the time of AVHRR measurements (degrees). Note that 1985, 1988, 2000, 2005, and 2009 contain AVHRR observations from the previous and the next NOAA satellites.

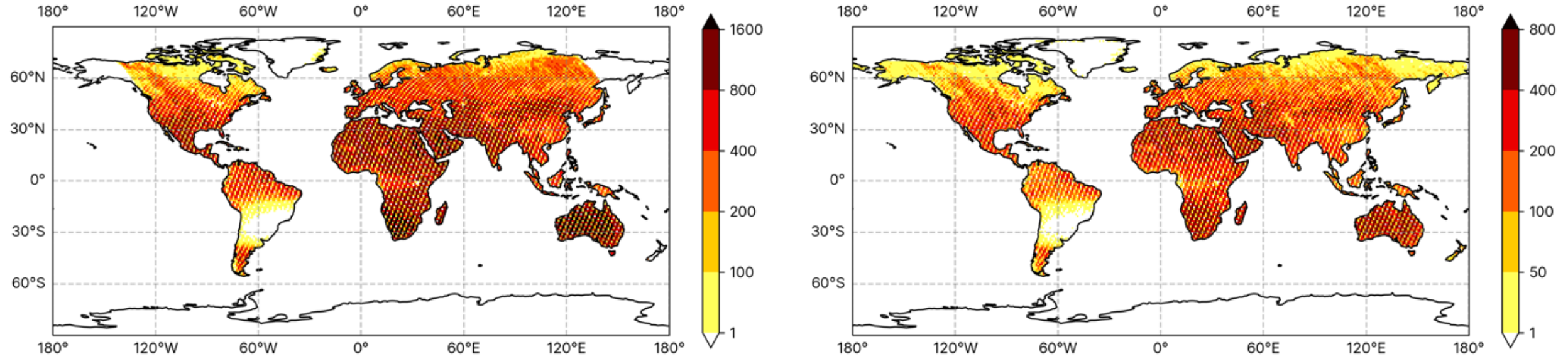

Figure S2: The spatial distributions of training (left) and test samples (right). Individual OCO-2 soundings are collocated with MODIS red and infrared reflectance bands and aggregated to 0.05° CMG to match the resolution of MCD43C1. Cells with fewer than 5 soundings are removed. The number of samples is counted at 1° resolution.

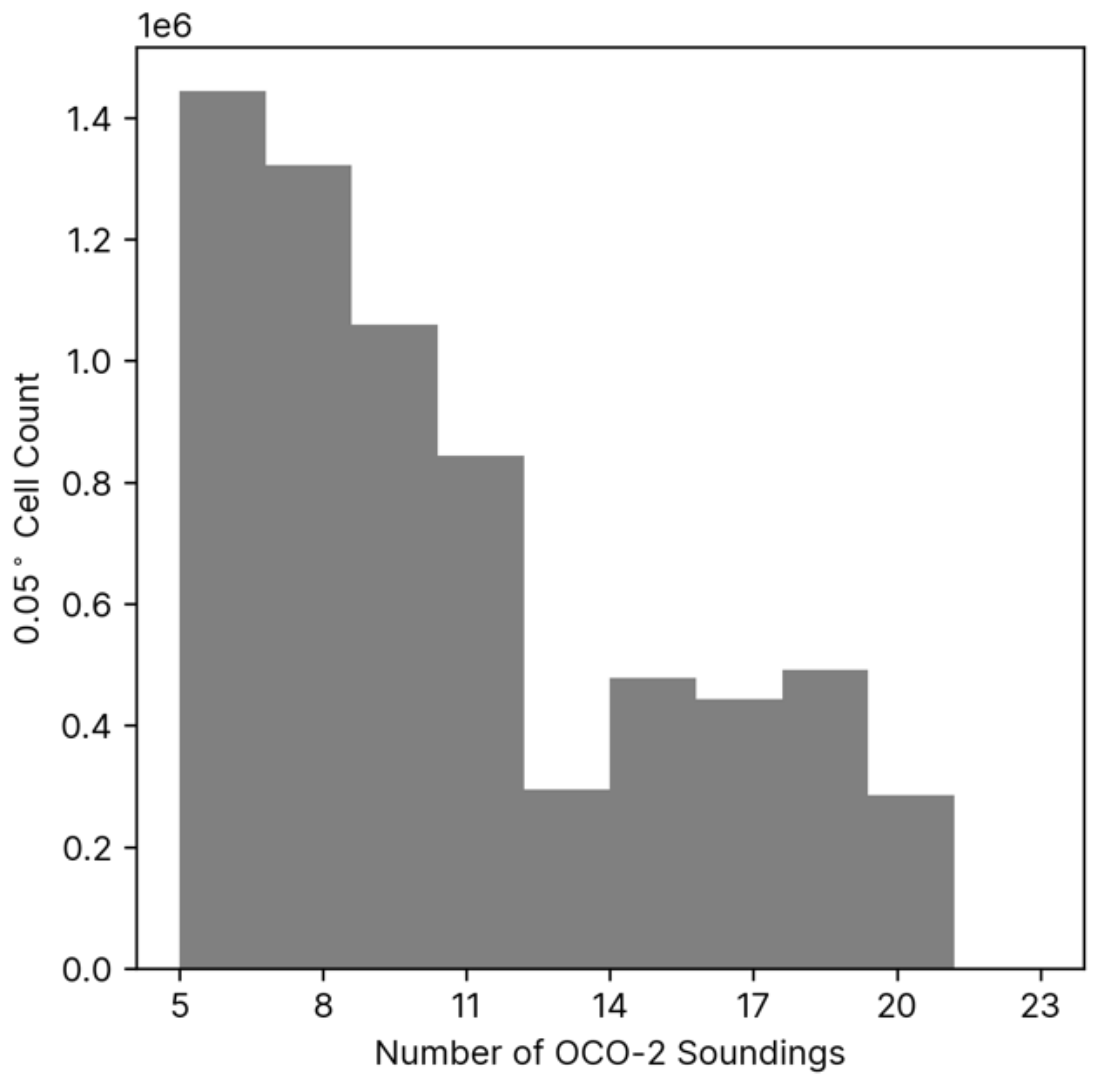

Figure S3: The distribution of OCO-2 sounding counts per 0.05° grid cell within the training dataset.

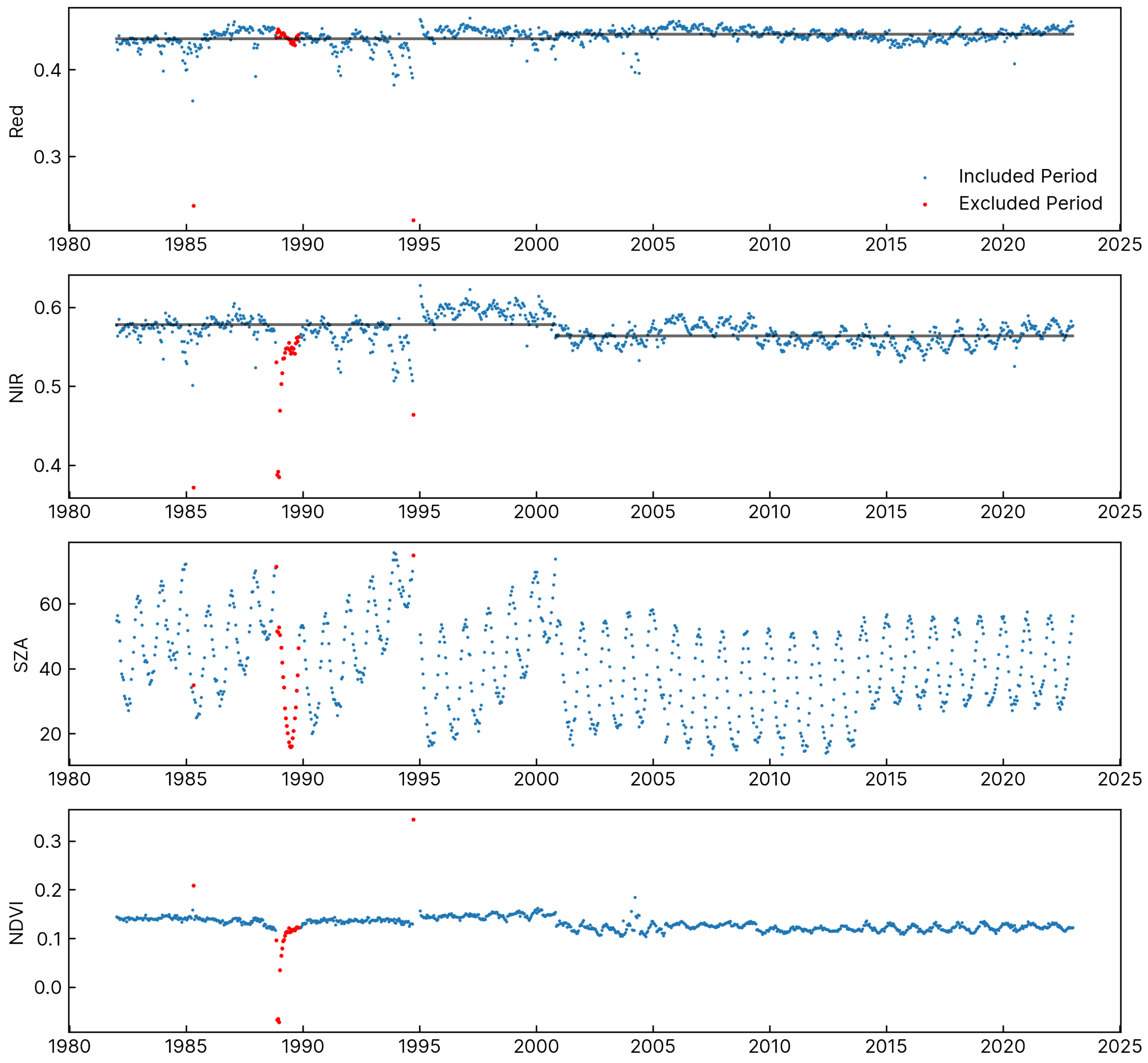

Figure S4. The upper two panels display the red and NIR reflectance of the LTDR AVHRR dataset spatially averaged over 10 calibration PICSs. A 5×5-pixel region centered on the coordinate of each PICSs is used to extract the reflectance. The black horizontal lines indicate the temporal mean between the AVHRR-2 and AVHRR-3 periods. The SZA and NDVI time series averaged across the same set of PICSs are shown in the lower two panels. Data points in red are anomalous periods excluded during the calibration (see Text S1 for justification). We later filled the excluded periods with the seasonal mean of the previous and following years immediately before the HANTS gap-filling reconstruct the complete time series.

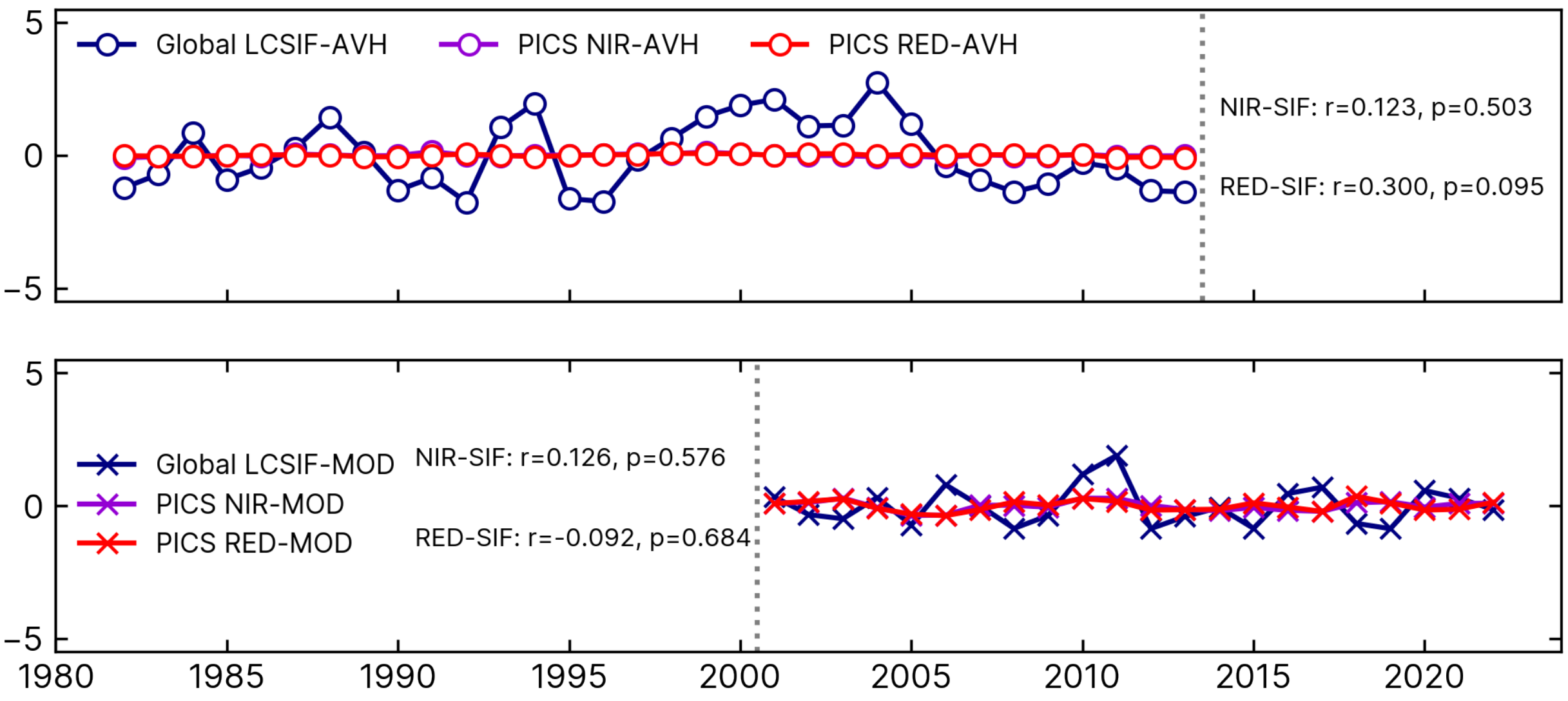

Figure S5: The upper panel shows the median of the annual time series of detrended normalized anomalies (%) of the calibrated red and near-infrared reflectance at the 10 validation PICSs and that of AVHRR-based global growing season mean LCSIF. Values are shown for 1982-2014 when there are expected solar zenith angle drifts within the LTDR dataset. The lower panel shows the median of the detrended normalized anomalies of the MODIS-based reflectance at PICSs and the global annual mean of LCSIF reconstructed from MODIS reflectance.

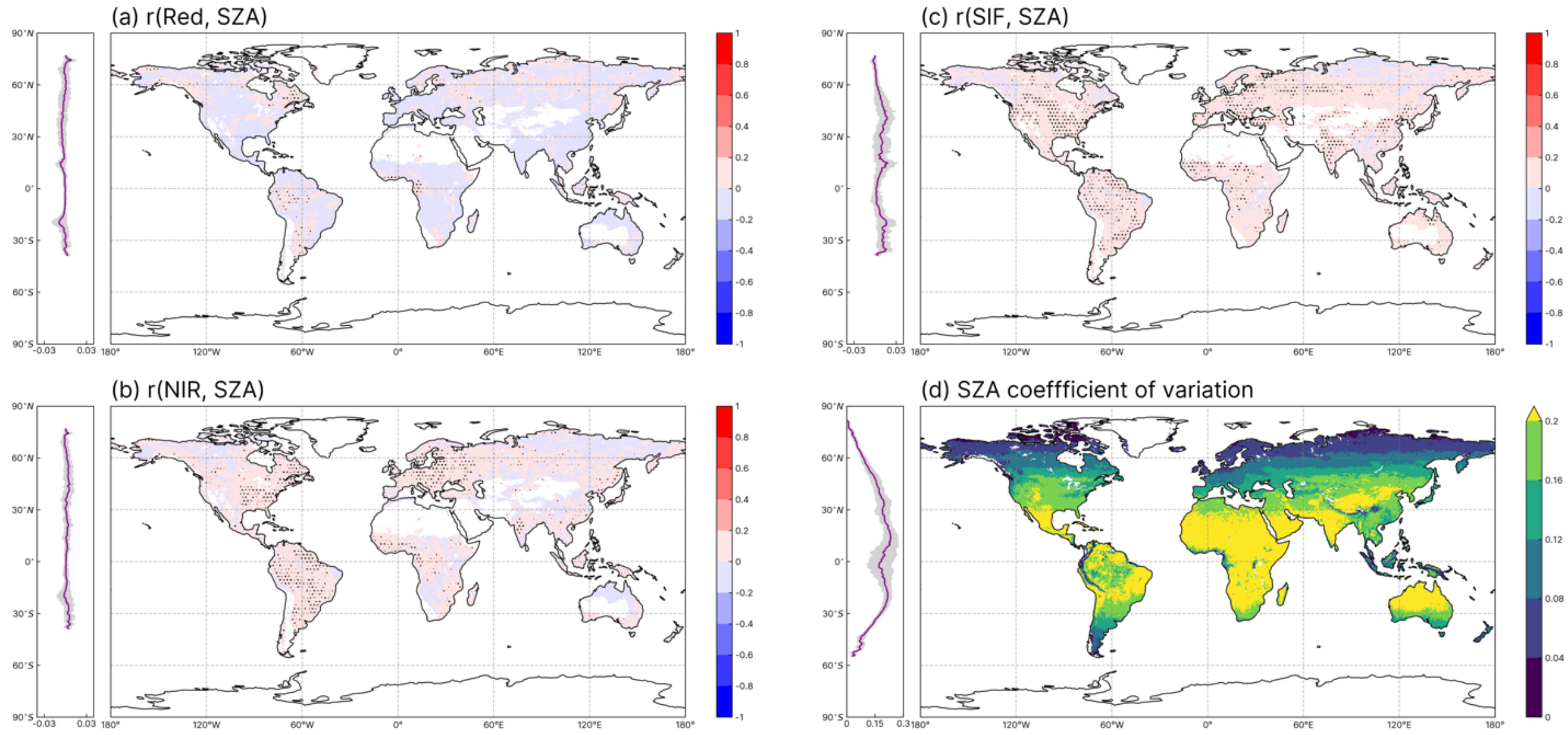

Figure S6. The remaining correlation between the solar zenith angle and the corrected (a) red and (b) NIR channel reflectance and (c) reconstructed LCSIF during the AVHRR period. The shaded area represents regions where the correlation is significant at a $p<0.05$ level. (d) The coefficient

of variation (standard deviation/mean) of annual mean SZA at the time of AVHRR measurements.

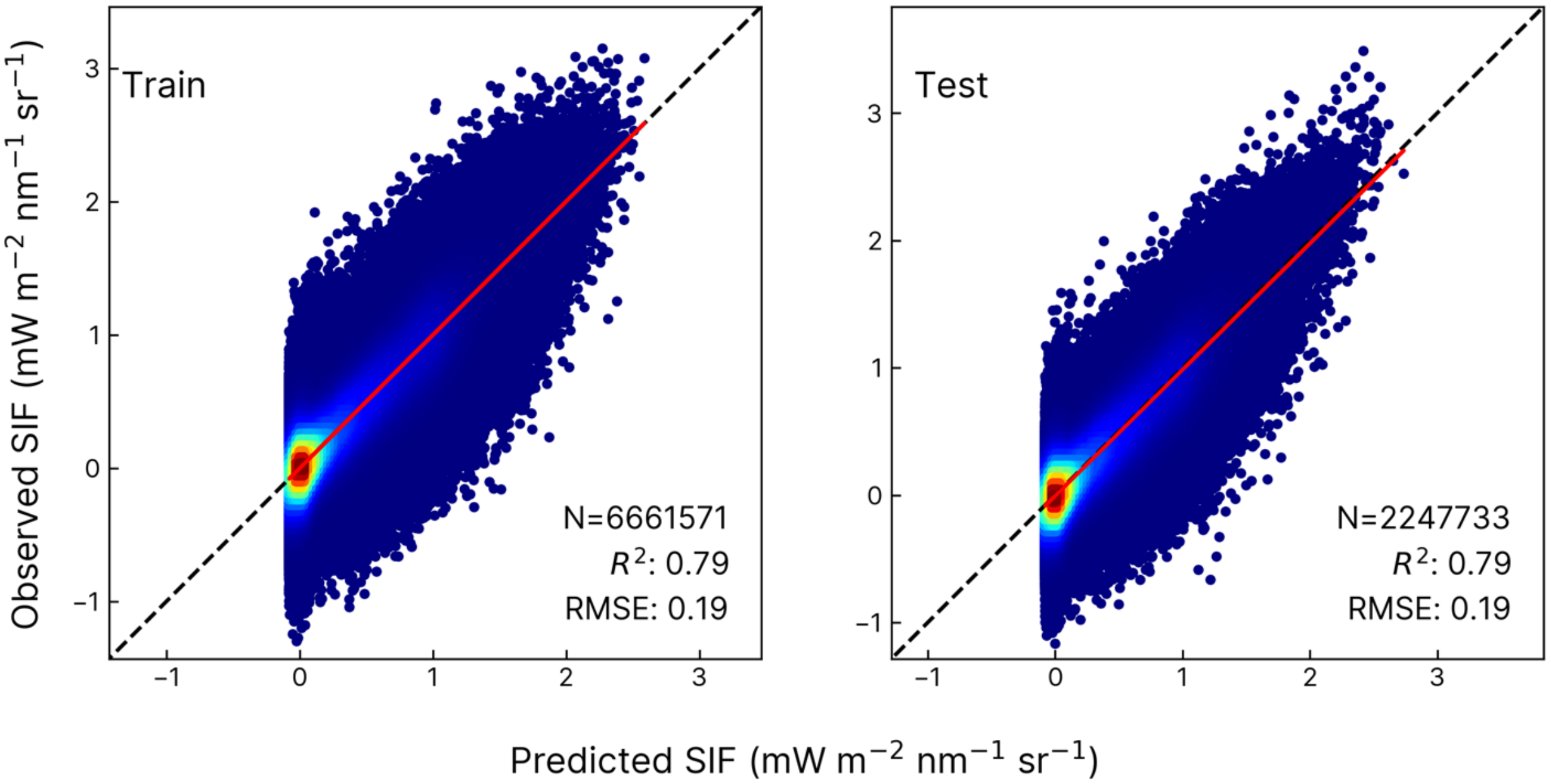

Figure S7. Predicted vs. observed SIF for training and test datasets, for all land cover types combined.

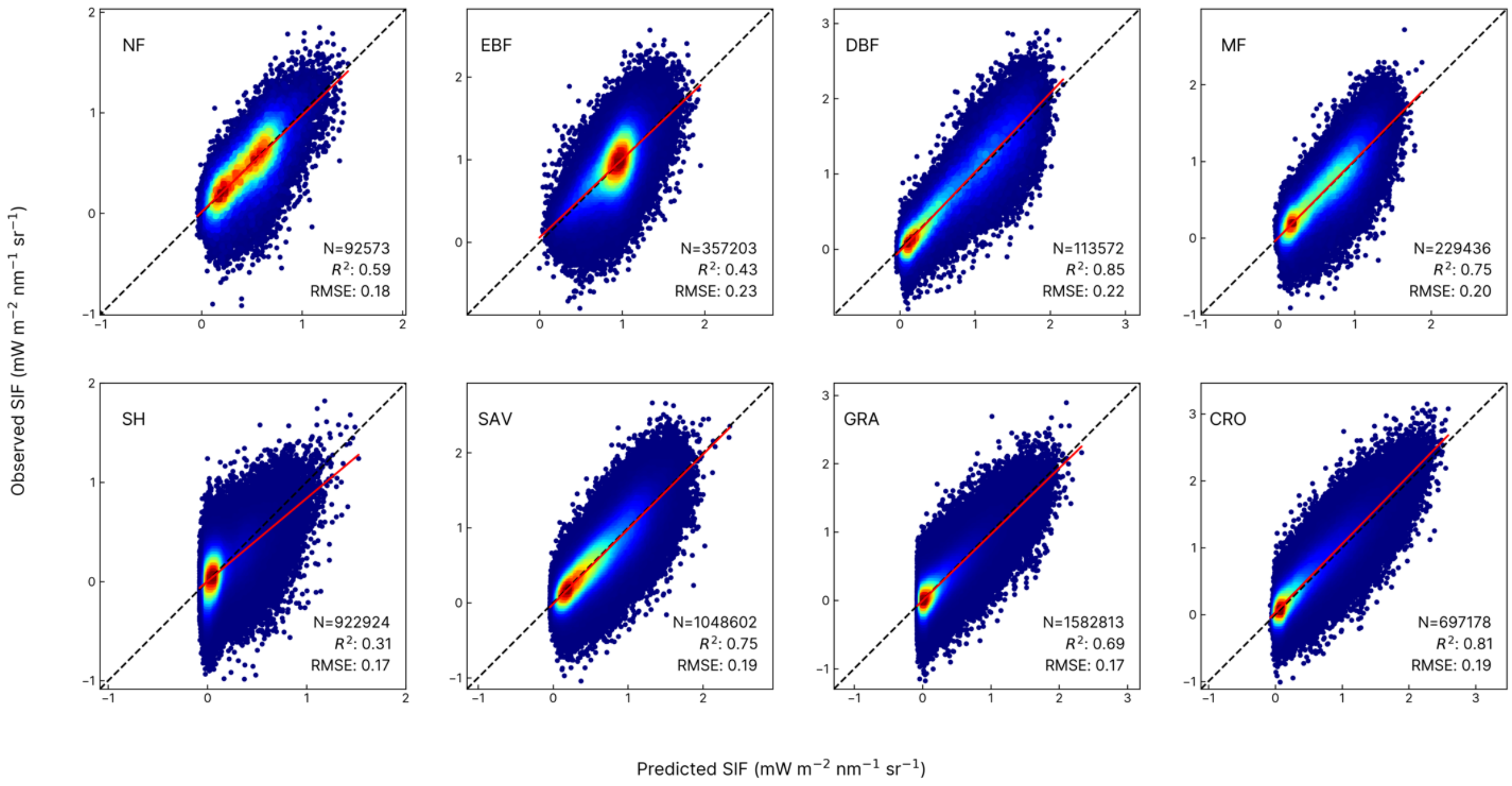

Figure S8. Same as Figure 3 but for the training dataset.

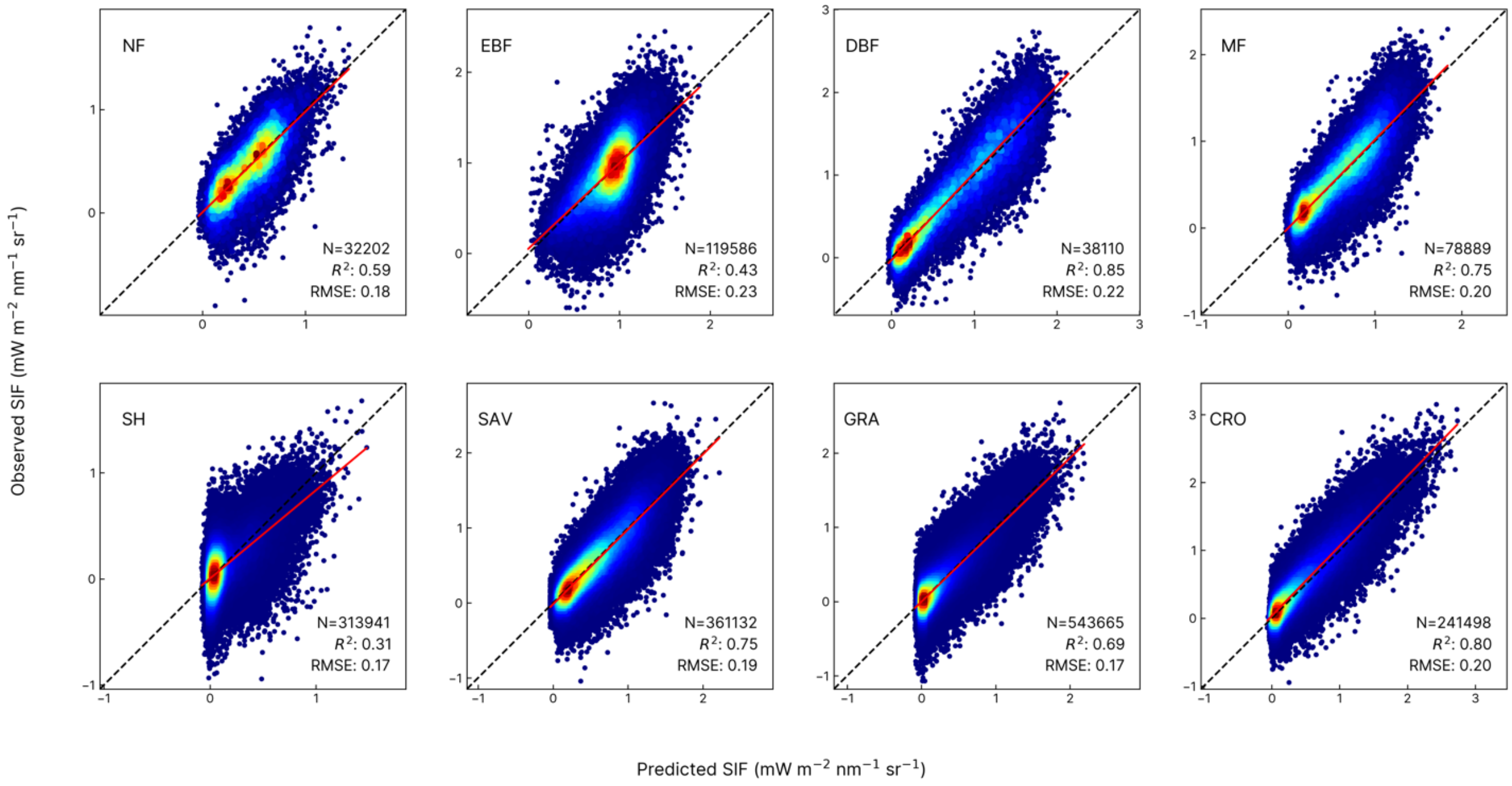

Figure S9. To examine whether the SIF model performance is sensitive to how we split the data into training and testing sets, we performed a sensitivity analysis by using data from 2015, 2017, 2018, 2019, 2021, and 2022 for training, and the data from 2016 and 2022 for testing. The resulting model performance is consistent with the original data split.

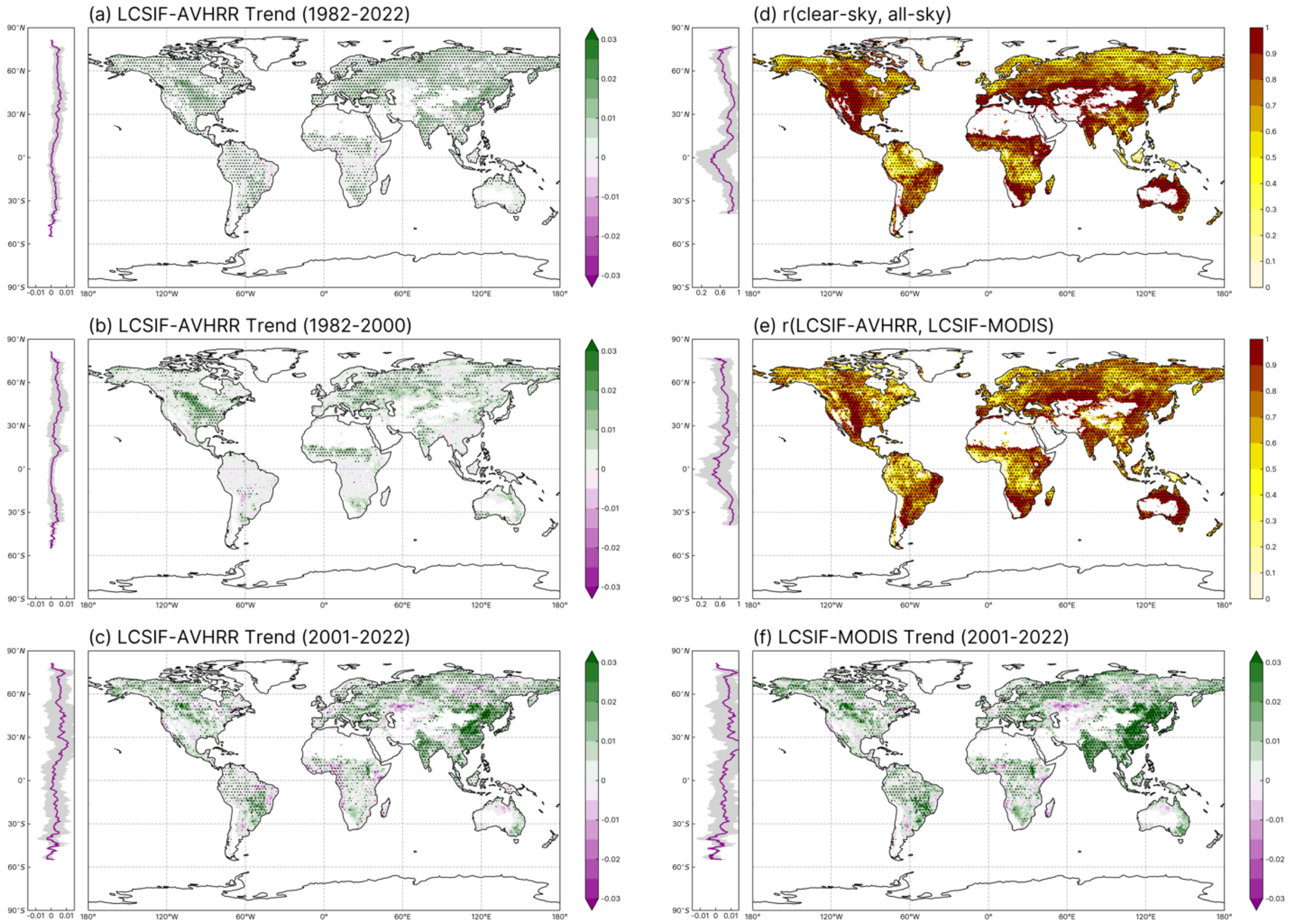

Figure S10: The spatial trends of AVHRR-based and MODIS-based LCSIF, same as Figure 5 in the main text but with trends computed using Theil-Sen slope estimator and significance determined by Hamed and Rao modified Mann-Kendall test with lag 1 autocorrelation and $\alpha=0.05$.

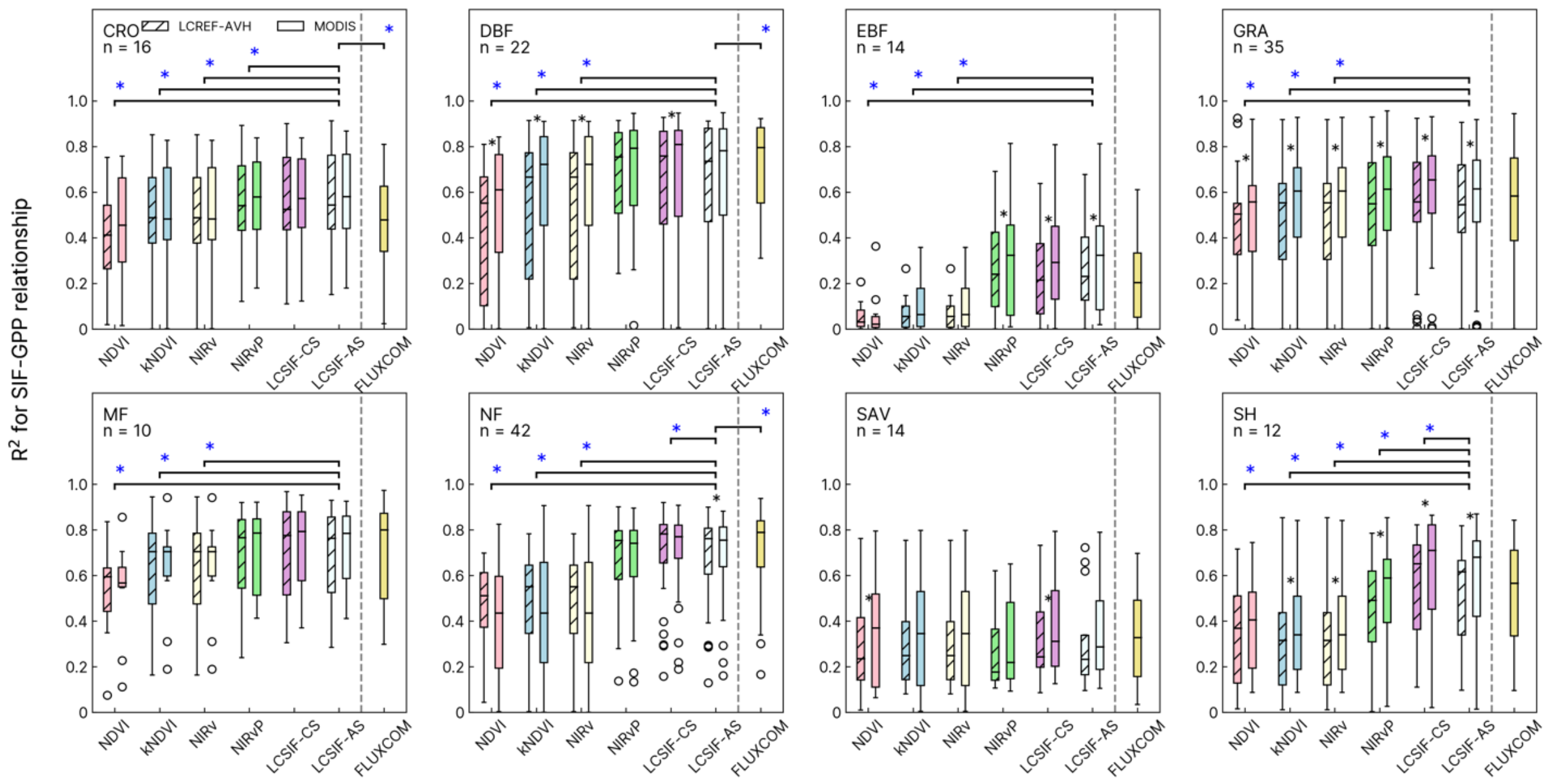

Figure S11: Comparing the strengths of the calibrated AVHRR-based (LCREF-AVH) or MODIS-based VI-GPP and SIF-GPP correlations at eddy-covariance sites. We used nighttime partitioned GPP from the FLUXNET 2015 dataset since 2001 for this analysis. The blue asterisk indicates whether the of the correlation with GPP is higher in the case of all-sky LCSIF (LCSIF-AS) than in a reflectance-only VI (paired t-test, one-tailed $p$<0.05), or whether there is a difference in the correlation with FLUXNET GPP between all-sky LCSIF (LCSIF-AS), FLUXCOM, NIRvP, or clear-sky LCSIF (LCSIF-CS) (paired t-test, two-tailed $p$<0.05). The black asterisk indicates for each type of VI or SIF variable, whether the MODIS-based version has a higher correlation with site-level GPP than the AVHRR-based product after calibration (one-tailed $p$<0.05).

Table S1: A comparison of recent reconstructed global SIF products

| Reconstructed SIF Product | Reconstruction Algorithm | Predictors | Target SIF Product | Period Available | Temp. & Spatial Resolution |
|---|---|---|---|---|---|
| LCSIF (this study) | ANN | MCD43C4: Band 1-2 | OCO-2 SIF (757nm) | 1982-2021 | 0.05° biweekly |
| RSIF(Gentine and Alemohammad, 2018) | ANN | MYD09A1: Band 1-4 | GOME-2 SIF (740nm) | 2007-2017 | 0.5° biweekly |
| CSIF(Zhang et al., 2018a) | ANN | MCD43C4: Band 1-4 | OCO-2 SIF (757nm) | 2000-2022 | 0.05° 4day |
| GOSIF(Li and Xiao, 2019) | Cubist regression tree | MCD43C4: EVI<br>MERRA-2: PAR, VPD Ta<br>MCD12C1: Land cover types | OCO-2 SIF (757nm) | 2000-2021 | 0.05° 8day |
| RTSIF (Chen et al., 2022) | XGboost | MCD43C4: Band 1-7, CERES PAR<br>MOD11C1: LST<br>MCD12C1: IGBP<br>ISLSCP II: C4 Percentage | TROPOMI SIF (740nm) | 2001-2020 | 0.05° 8day |
| SDSIF(Hu et al., 2022) | ANN | MCD43C4: Band 1-4 NDWI<br>ERA5: VPD, Ta | TROPOMI SIF (743-758 nm) | 2018-2020 | 0.05° daily & 4day |
| LUE-SIF(Duveiller et al., 2020) | Semi-empirical LUE model | MCD43C4: NIRv, NDWI<br>LST: MYD11C2 | GOME-2 and SCIAMACHY SIF (740nm) | 2007-2018 | 0.05° 8day |
| HSIF(Wen et al., 2020) | Random Forest with regionalization constraints + CDF matching harmonization | MCD43C4: Band 1-7 | GOME-2 and SCIAMACHY SIF (740nm) | 2008-2018 | 0.05° monthly |
| ST-LGBM SIF (Shen et al., 2022) | Spatiotemporal constrained light gradient boosting (ST-LGBM) | MCD43C4: NIRv<br>MERRA-2: PAR, VPD, Ta<br>MCD12C1: LC types<br>OCO-2: Temporal and Spatial Factors | OCO-2 SIF (757nm) | Unknown | 0.05° 8day |

Table S2: The sensor-origin of LTDR AVHRR data used in this study

| AVHRR Platform | Begin Date (YYYYDOY) | End Date (YYYYDOY) |
|---|---|---|
| N07 | 1982001 | 1985003 |
| N09 | 1985004 | 1988312 |
| N11 | 1988313 | 1994365 |
| N14 | 1995001 | 2000305 |
| N16 | 2000306 | 2005181 |
| N18 | 2005182 | 2009151 |
| N19 | 2009152 | 2013365 |
| M1 | 2014001 | 2023181 |

Table S3. A list of PICSs used for evaluating residual orbital effects in AVHRR

| Site | Latitude | Longitude |
|---|---|---|
| Arabia2 | 20.19 | 51.63 |
| Sudan1* | 22.11 | 28.11 |
| Arabia1* | 19.80 | 47.07 |
| Egypt1 | 26.61 | 26.22 |
| Libya3* | 23.22 | 23.23 |
| Libya2 | 25.08 | 20.77 |
| Algeria3 | 30.63 | 7.83 |
| Mauritania1* | 19.51 | -8.57 |
| Mali1 | 19.14 | -5.77 |
| Libya4 | 28.67 | 23.42 |
| Niger1 | 20.26 | 9.64 |
| Algeria1 | 23.83 | -0.76 |
| Mauritania2* | 19.78 | -8.89 |
| Algeria4* | 29.99 | 5.10 |
| Libya1* | 24.65 | 13.25 |
| Algeria5* | 31.16 | 2.24 |
| Algeria2* | 25.99 | -0.62 |
| Niger2* | 21.33 | 10.60 |
| Niger3 | 21.51 | 7.86 |
| Arabia3 | 28.80 | 43.05 |

* 10 PICSs randomly selected for validation.

Table S4: List of FLUXNET sites used for evaluating LCSIF and VIs

| SITE_ID | LAT | LON | IGBP |
|---|---|---|---|
| AR-SLu | -33.4648 | -66.4598 | MF |
| AR-Vir | -28.2395 | -56.1886 | ENF |
| AT-Neu | 47.11667 | 11.3175 | GRA |
| AU-Ade | -13.0769 | 131.1178 | WSA |
| AU-ASM | -22.283 | 133.249 | SAV |
| AU-Cpr | -34.0021 | 140.5891 | SAV |
| AU-Cum | -33.61518 | 150.72362 | EBF |
| AU-DaP | -14.0633 | 131.3181 | GRA |
| AU-DaS | -14.1593 | 131.3881 | SAV |
| AU-Dry | -15.2588 | 132.3706 | SAV |
| AU-Emr | -23.8587 | 148.4746 | GRA |
| AU-Gin | -31.3764 | 115.7138 | WSA |
| AU-GWW | -30.1913 | 120.6541 | SAV |
| AU-How | -12.4943 | 131.1523 | WSA |
| AU-Lox | -34.4704 | 140.6551 | DBF |
| AU-RDF | -14.5636 | 132.4776 | WSA |
| AU-Rig | -36.6499 | 145.5759 | GRA |
| AU-Stp | -17.1507 | 133.3502 | GRA |
| AU-TTE | -22.287 | 133.64 | GRA |
| AU-Tum | -35.6566 | 148.1517 | EBF |
| AU-Wac | -37.4259 | 145.1878 | EBF |
| AU-Whr | -36.6732 | 145.0294 | EBF |
| AU-Wom | -37.4222 | 144.0944 | EBF |
| AU-Ync | -34.9893 | 146.2907 | GRA |
| BE-Bra* | 51.30761 | 4.51984 | MF |
| BE-Lon | 50.55162 | 4.74623 | CRO |
| BE-Vie* | 50.30493 | 5.99812 | MF |
| BR-Sa1 | -2.85667 | -54.95889 | EBF |
| BR-Sa3 | -3.01803 | -54.97144 | EBF |
| CA-Gro | 48.2167 | -82.1556 | MF |
| CA-Man* | 55.87962 | -98.48081 | ENF |
| CA-NS1 | 55.87917 | -98.48389 | ENF |
| CA-NS2 | 55.90583 | -98.52472 | ENF |
| CA-NS3 | 55.91167 | -98.38222 | ENF |
| CA-NS4 | 55.91437 | -98.380645 | ENF |
| CA-NS5 | 55.86306 | -98.485 | ENF |
| CA-NS6 | 55.91667 | -98.96444 | OSH |
| CA-NS7 | 56.63583 | -99.94833 | OSH |
| CA-Oas* | 53.62889 | -106.19779 | DBF |
| CA-Obs* | 53.98717 | -105.11779 | ENF |
| CA-Qfo | 49.6925 | -74.34206 | ENF |
| CA-SF1 | 54.48503 | -105.81757 | ENF |
| CA-SF2 | 54.25392 | -105.8775 | ENF |
| CA-SF3 | 54.09156 | -106.00526 | OSH |
| CA-TP1 | 42.6609361 | -80.559519 | ENF |
| CA-TP2 | 42.7744194 | -80.458775 | ENF |
| CA-TP3 | 42.7068111 | -80.348314 | ENF |
| CA-TP4 | 42.710161 | -80.357376 | ENF |
| CA-TPD | 42.635328 | -80.557731 | DBF |
| CG-Tch | -4.28917 | 11.65642 | SAV |
| CH-Cha | 47.21022 | 8.41044 | GRA |
| CH-Dav* | 46.81533 | 9.85591 | ENF |
| CH-Fru | 47.11583 | 8.53778 | GRA |
| CH-Lae | 47.47833 | 8.36439 | MF |
| CH-Oe1 | 47.28583 | 7.73194 | GRA |
| CH-Oe2 | 47.28642 | 7.73375 | CRO |
| CN-Cha | 42.4025 | 128.0958 | MF |
| CN-Cng | 44.5934 | 123.5092 | GRA |
| CN-Dan | 30.4978 | 91.0664 | GRA |
| CN-Din | 23.1733 | 112.5361 | EBF |
| CN-Du2 | 42.0467 | 116.2836 | GRA |
| CN-Du3 | 42.0551 | 116.2809 | GRA |
| CN-HaM | 37.37 | 101.18 | GRA |
| CN-Qia | 26.7414 | 115.0581 | ENF |
| CN-Sw2 | 41.7902 | 111.8971 | GRA |
| CZ-BK1 | 49.50208 | 18.53688 | ENF |
| CZ-BK2 | 49.49443 | 18.54285 | GRA |
| DE-Geb | 51.09973 | 10.91463 | CRO |
| DE-Hai | 51.07921 | 10.45217 | DBF |
| DE-Kli | 50.89306 | 13.52238 | CRO |
| DE-Lkb | 49.09962 | 13.30467 | ENF |
| DE-Lnf | 51.32822 | 10.3678 | DBF |
| DE-Obe | 50.78666 | 13.72129 | ENF |
| DE-RuR | 50.62191 | 6.30413 | GRA |
| DE-Seh | 50.87062 | 6.44965 | CRO |
| DE-Tha* | 50.96256 | 13.56515 | ENF |
| DK-Eng | 55.69053 | 12.19175 | GRA |
| DK-Sor* | 55.48587 | 11.64464 | DBF |
| ES-Amo | 36.83361 | -2.25232 | OSH |
| ES-LgS | 37.09794 | -2.96583 | OSH |
| ES-LJu | 36.92659 | -2.75212 | OSH |
| FI-Hyy* | 61.84741 | 24.29477 | ENF |
| FI-Jok | 60.8986 | 23.51345 | CRO |
| FI-Let | 60.64183 | 23.95952 | ENF |
| FI-Sod | 67.36239 | 26.63859 | ENF |
| FR-Fon | 48.47636 | 2.7801 | DBF |
| FR-Gri | 48.84422 | 1.95191 | CRO |
| FR-LBr* | 44.71711 | -0.7693 | ENF |
| FR-Pue | 43.7413 | 3.5957 | EBF |
| GF-Guy | 5.27877 | -52.92486 | EBF |
| GH-Ank | 5.26854 | -2.69421 | EBF |
| GL-ZaH | 74.47328 | -20.5503 | GRA |
| IT-BCi | 40.52375 | 14.95744 | CRO |
| IT-CA1 | 42.38041 | 12.02656 | DBF |
| IT-CA2 | 42.37722 | 12.02604 | CRO |
| IT-CA3 | 42.38 | 12.0222 | DBF |
| IT-Col* | 41.84936 | 13.58814 | DBF |
| IT-Cp2 | 41.70427 | 12.35729 | EBF |
| IT-Cpz* | 41.70525 | 12.37611 | EBF |
| IT-Isp | 45.81264 | 8.63358 | DBF |
| IT-La2 | 45.9542 | 11.2853 | ENF |
| IT-Lav | 45.9562 | 11.28132 | ENF |
| IT-MBo | 46.01468 | 11.04583 | GRA |
| IT-Noe | 40.60618 | 8.15169 | CSH |
| IT-PT1 | 45.20087 | 9.06104 | DBF |
| IT-Ren* | 46.58686 | 11.43369 | ENF |
| IT-Ro1 | 42.40812 | 11.93001 | DBF |
| IT-Ro2 | 42.39026 | 11.92093 | DBF |
| IT-Tor | 45.84444 | 7.57806 | GRA |
| JP-MBF | 44.3869 | 142.3186 | DBF |
| JP-SMF | 35.2617 | 137.0788 | MF |
| MY-PSO | 2.973 | 102.3062 | EBF |
| NL-Hor | 52.24035 | 5.0713 | GRA |
| NL-Loo* | 52.16658 | 5.74356 | ENF |
| PA-SPn | 9.31814 | -79.6346 | DBF |
| PA-SPs | 9.31378 | -79.63143 | GRA |
| RU-Cok | 70.82914 | 147.49428 | OSH |
| RU-Fyo* | 56.46153 | 32.92208 | ENF |
| RU-Ha1 | 54.72517 | 90.00215 | GRA |
| SD-Dem | 13.2829 | 30.4783 | SAV |
| SJ-Blv | 78.92163 | 11.83109 | SNO |
| SN-Dhr | 15.40278 | -15.43222 | SAV |
| US-AR1 | 36.4267 | -99.42 | GRA |
| US-AR2 | 36.6358 | -99.5975 | GRA |
| US-ARb | 35.5497 | -98.0402 | GRA |
| US-ARc | 35.54649 | -98.04 | GRA |
| US-ARM | 36.6058 | -97.4888 | CRO |
| US-Blo* | 38.8953 | -120.6328 | ENF |
| US-Cop | 38.09 | -109.39 | GRA |
| US-CRT | 41.628495 | -83.347086 | CRO |
| US-GBT | 41.36579 | -106.2397 | ENF |
| US-GLE | 41.36653 | -106.2399 | ENF |
| US-Goo | 34.2547 | -89.8735 | GRA |
| US-Ha1* | 42.5378 | -72.1715 | DBF |
| US-IB2 | 41.84062 | -88.24103 | GRA |
| US-KS2 | 28.6086 | -80.6715 | CSH |
| US-Lin | 36.3566 | -119.8423 | CRO |
| US-Me1 | 44.5794 | -121.5 | ENF |
| US-Me2 | 44.4523 | -121.5574 | ENF |
| US-Me3 | 44.3154 | -121.6078 | ENF |
| US-Me5 | 44.43719 | -121.56676 | ENF |
| US-Me6 | 44.3232842 | -121.6078 | ENF |
| US-MMS | 39.3232 | -86.4131 | DBF |
| US-Ne1 | 41.16506 | -96.47664 | CRO |
| US-Ne2 | 41.16487 | -96.4701 | CRO |
| US-Ne3 | 41.17967 | -96.43965 | CRO |
| US-NR1 | 40.0329 | -105.5464 | ENF |
| US-Oho | 41.5545 | -83.8438 | DBF |
| US-Pfa* | 45.9459 | -90.2723 | MF |
| US-Prr | 65.12367 | -147.48756 | ENF |
| US-SRC | 31.9083 | -110.8395 | MF |
| US-SRG | 31.789379 | -110.82768 | GRA |
| US-SRM | 31.8214 | -110.8661 | WSA |
| US-Sta | 41.3966 | -106.8024 | OSH |
| US-Syv | 46.242 | -89.3477 | MF |
| US-Ton | 38.4316 | -120.96598 | WSA |
| US-Tw2 | 38.1047 | -121.6433 | CRO |
| US-Var | 38.4133 | -120.9507 | GRA |
| US-WCr | 45.8059 | -90.0799 | DBF |
| US-Whs | 31.7438 | -110.0522 | OSH |
| US-Wi3 | 46.634722 | -91.098667 | DBF |
| US-Wi4 | 46.739333 | -91.16625 | ENF |
| US-Wi6 | 46.624889 | -91.298222 | OSH |
| US-Wi9 | 46.618778 | -91.081444 | ENF |
| US-Wkg | 31.7365 | -109.9419 | GRA |
| ZM-Mon | -15.4391 | 23.2525 | DBF |

* Sites with at least three years of data in both MODIS and AVHRR periods.

Text S1 Details about anomalous data removal in the AVHRR LTDR record

We excluded three periods from the AVHRR LTDR record during prescreening. We first removed the second half of April, 1985 when we identified a sharp NDVI anomaly over 10 calibration PICSs that we deemed a short-term artifact. The second excluded period (November 1988-October 1989) was a documented low-quality data period at the beginning of the N11(Tian et al., 2015). The third period removed (September 16, 1994-December 31, 1994) was due to a scan motor error of the N11 AVHRR sensor on September 13, 1994, causing subsequent data gaps until the end of the N11 mission (https://www.ospo.noaa.gov/Operations/POES/NOAA11/avhrr.html).

Text S2 Additional details for the machine learning model used for AVHRR-MODIS spatial residual correction

As briefly described in Section 2.3 of the main text, we employed both local and global correlational structures between overlapping MODIS and AVHRR observations for calibration. After correcting the biases between AVHRR and MODIS using a local linear model, we calculated the difference between MODIS observations and linearly-corrected AVHRR values as $\gamma_{residual}$, Equation 8 in the main text). We assumed $\gamma_{residual}$ could originate from the non-linear responses of reflectance retrieved by MODIS and AVHRR instruments to environmental variates such as atmospheric (aerosol and cloud cover) and surface conditions (topography, snow, and soil background noises). To capture the effect of these environmental covariates, we randomly subsampled 2 million samples of the global $\gamma_{residual}$ values from 2001-2022 to form a training dataset (except for 2002, 2010, 2018, which were reserved for validation). The neural network regression predictors are standardized aerosol optical depth, snow depth, cloud cover, elevation, and the linearly corrected AVHRR values. We used a feed-forward neural network with 3 hidden layers and 64 neurons. The model was trained for 10 epochs with a minibatch size 1024 at a learning rate of 0.001. We then added the predicted $\gamma_{residual}$ values to the linearly corrected AVHRR values for the entire AVHRR record.